\documentclass[12pt,preprint]{aastex}

\newcommand{\myemail}{eliad@le.infn.it}
\newcommand{\CO}{$^{12}$CO(1-0)}
\newcommand{\COO}{$^{13}$CO(2-1)}

\shorttitle{Mapping Vela Molecular Ridge Cloud D}
\shortauthors{Elia et al.}

\begin{document}

\title{Mapping molecular emission in Vela Molecular Ridge Cloud D}

\author{D. Elia\altaffilmark{1}, F. Massi\altaffilmark{2}, F. Strafella\altaffilmark{1,3},
M. De Luca\altaffilmark{4,5}, T. Giannini\altaffilmark{5}, D.
Lorenzetti\altaffilmark{5}, B. Nisini\altaffilmark{5}, L.
Campeggio\altaffilmark{1} and B. M. T. Maiolo\altaffilmark{1}}

\altaffiltext{1}{Dipartimento di Fisica, Universit\`a di Lecce, CP
193, I-73100 Lecce, Italy} \email{\myemail} \altaffiltext{2}{INAF -
Osservatorio Astrofisico di Arcetri, Largo E. Fermi 5, I-50125
Firenze, Italy} \altaffiltext{3}{INAF - Istituto di Fisica dello
Spazio Interplanetario, Via Fosso del Cavaliere 100, I-00133 Roma,
Italy} \altaffiltext{4}{Dipartimento di Fisica, Universit\`{a} degli
Studi di Roma "Tor Vergata", Via della Ricerca Scientifica 1,
I-00133 Roma, Italy}
 \altaffiltext{5}{INAF - Osservatorio Astronomico di Roma, Via
Frascati 33, I-00040 Monteporzio Catone, Italy}

\begin{abstract}
We present the \CO~and \COO~line maps obtained observing with the
SEST a $\sim 1\degr \times 1\degr$ region of the Vela Molecular
Ridge, Cloud~D. This cloud is part of an intermediate-mass star
forming region that is relatively close to the Sun. Our observations
reveal, over a wide range of spatial scales (from $\sim 0.1$ to a
few parsecs), a variety of dense structures such as arcs, filaments
and clumps, that are in many cases associated with far-IR point-like
sources, recognized as young stellar objects and embedded star
clusters. The velocity field analysis highlights the presence of
possible expanding shells, extending over several parsecs, probably
related to the star forming activity of the cloud. Furthermore, the
analysis of the line shapes in the vicinity of the far-IR sources
allowed the detection of 13 molecular outflows. Considering a
hierarchical scenario for the gas structure, a cloud decomposition
was obtained for both spectral lines by means of the CLUMPFIND
algorithm. The CLUMPFIND output has been discussed critically and a
method is proposed to reasonably correct the list of the identified
clumps. We find that the corresponding mass spectrum shows a
spectral index $\alpha\sim1.3 \div 2.0$ and the derived clump masses
are below the corresponding virial masses. The mass-radius and
velocity dispersion-radius relationships are also briefly discussed
for the recovered clump population.

\end{abstract}


\section{Introduction}

Vela Molecular Ridge is a complex located in the southern sky,
extending in galactic longitudes $\ell=260\degr~\div~273\degr$, and
confined to latitudes $b=\pm~2\degr$, as reported by \citet{may88},
who carried out an extensive survey of the third galactic quadrant
($\ell=210\degr \div 279\degr$, $b=\pm5\degr$) in the
\CO~transition, with a spatial resolution of $0.5\degr$. This
complex is composed of at least three molecular clouds, named A, C
and~D \citep{mur91}, at a distance of $d=700\pm 200$~pc
\citep{lis92}, and a more distant cloud, named~B, at $d \sim
2000$~pc. All these regions (hereinafter, VMR-A, -B, -C, -D,
respectively) contain a gas mass exceeding a total of
$10^5$~M$_{\sun}$ and are active sites of star formation, as
revealed by the \emph{IRAS} mission \citep{lis92}.

VMR-D is one of the nearest massive star forming regions and
therefore one of the most studied. A detailed analysis aimed to
investigate and describe the star formation activity in this cloud
has been carried out in the last decade through multifrequency
observations \citep{lor93,mas99,mas00,gia05,mas06a}, that have
pointed out the evidence for young stars clustering.

\defcitealias{wou99}{WB99}
At radio wavelengths, further observations of VMR-D have been
carried out by \citet[hereinafter WB99]{wou99} who mapped, in \CO,
$^{13}$CO(1-0), C$^{18}$O(1-0) and CS(2-1) transitions, nine zones
towards bright infrared sources in VMR field including IRS~17,
IRS~19, and IRS~20 \citep[here we shall use their nomenclature for
the protostellar \emph{IRAS} sources in common with their
sample]{lis92}. In particular, in the case of the three mentioned
sources, the association with a gas clump and a molecular outflow
has also been ascertained.

\defcitealias{yam99}{YA99}
\citet[hereinafter YA99]{yam99} obtained large-scale \CO~and
$^{13}$CO(1-0) maps of the whole VMR with NANTEN telescope, with a
grid spacing of 8$\arcmin$. They also mapped the densest regions in
the C$^{18}$O(1-0) line with a grid spacing of 2$\arcmin$. The total
mass of the complex estimated by means of these three molecular
tracers is $M_{gas}$($^{12}$CO)$ \sim5.6\times 10^{5}$~M$_{\sun}$,
$M_{gas}$($^{13}$CO)$ \sim1.4\times 10^{5}$~M$_{\sun}$, and
$M_{gas}$(C$^{18}$O)$ \sim5.3\times 10^{4}$~M$_{\sun}$,
respectively. Furthermore, these authors also related the gas
distribution to the position of the IR sources in the field, in
order to investigate the nature of the star formation in this
complex.

As a further progress in the knowledge of the VMR gas morphology,
\citet{mor01} repeated the \CO~observations of the ridge using the
same instrumentation of \citetalias{yam99}, but mapping it with a
finer grid spacing of 2$\arcmin$. These authors remark the highly
filamentary distribution of the gas emission and discuss the
possible interaction between VMR and the well-known Vela supernova
remnant (SNR), associated with a pulsar located at $\ell\sim
263.5\degr$, $b\sim -2.5\degr$. They conclude that the molecular gas
is associated to the SNR, but it is pre-existent.

More recently, \citet{fon05}, studying a conspicuous sample of
high-mass protostellar candidates in the southern sky, considered
also IRS~21 and found that this object is associated with CS(2-1),
CS(3-2), C$^{17}$O(1-0) lines and 1.2~mm continuum emission peaks.

\defcitealias{mas06b}{MA06}
Dust continuum millimeter observations of a $\sim 1\degr \times
1\degr$ portion of VMR-D (the same as investigated in this paper)
have been obtained with SIMBA/SEST \citep{mas05,del05,mas06b}
revealing both filamentary and clumpy appearance. In particular,
\citet{del05} compiled a list of 29 detected clumps \citep[discussed
in][hereinafter MA06]{mas06b} in most cases exhibiting non-gaussian
shapes and a tendency to clusterize, both suggesting a multiple star
formation activity.

In addition to this millimeter survey, further SIMBA observations
are available for the neighborhoods of IRS~17 \citep{fau04,gia05},
IRS~16, IRS~18, and IRS~20 \citep{bel06}, that allowed to detect
dust cores and determine their masses and luminosities.

The present work contributes to investigate the star formation
activity in a significant part of VMR-D by means of molecular line
observations down to spatial scales smaller than those investigated
in the preceding literature. This has been possible by exploiting
the spatial resolution of the SEST telescope that has been used to
observe this cloud in two millimetric bands corresponding to the
transitions \CO~and \COO, the first one suitable for tracing the
large-scale distribution of the gas, the second one useful for
better probing the highest column densities. In this way, several
aspects related to the star formation can be investigated, such as,
e.g., the gas distribution and kinematics to scales smaller than
1~pc (the typical size of pre-stellar cores), the correlation
between the intensity peaks of the maps and the position of the
brightest infrared point sources, the presence and census of
proto-stellar outflows in this region.

In this paper we focus our attention on the morphological and
kinematical aspects of the gas distribution. In Section~\ref{obs} we
provide a short description of our observations and of the data
reduction procedure. In Section~\ref{results}, after presenting the
integrated intensity and channel maps of the emission in both lines,
we discuss some significant velocity-position (vel-pos hereinafter)
diagrams, which highlight the complex structure of the gas velocity
field in this cloud. Section~\ref{clumps} illustrates the search for
clumps in the cloud, based on the CLUMPFIND code, and presents the
derived mass distribution. Finally, in Section~\ref{summary} a
summary of the main results is given.

\section{Observations and data reduction}\label{obs}
In the cloud identification by \citet{mur91}, VMR-D is located in
the rightmost end of the galactic longitude interval of the complex,
and shows two main CO emission peaks. The same characteristic is
recognizable, but in greater detail, in the \CO~and $^{13}$CO(1-0)
maps of \citetalias{yam99} and in the \CO~map of \citet{mor01}. Our
observations encompass the region of the higher longitude peak (the
smaller in size), located at $\ell\sim263\degr \div 265\degr$,
$b\sim0\div -1\degr$ \citep[see also Figure~1 in][]{mas03}.
Hereinafter, with VMR-D we shall indicate this part of cloud.

VMR-D was mapped in the \CO~($\nu=115.271$~GHz)
and~\COO~($\nu=220.399$~GHz) lines with the 15-m Swedish-ESO
telescope \citep[SEST, see][]{boo89} at La Silla, Chile, during two
complementary observational campaigns, in September 1999 and January
2003, respectively.

A high-resolution Acousto-Optical Spectrometer was used as a
backend, with a total bandwidth of about 100~MHz and a resolution of
41.7~kHz, in splitted mode, i.e.~in a configuration allowing the use
of both 115 and 230~GHz receivers simultaneously (1000 channels for
each receiver). This corresponds to velocity resolutions of $\sim
0.12$~km~s$^{-1}$ at 115~GHz and $\sim 0.06$~km~s$^{-1}$ at 220~GHz.
The FWHM of the primary beam and the main beam efficiency factors
are 45\arcsec~and 0.7 at 115~GHz, and 23\arcsec~and 0.5 at 220~GHz,
respectively. During the 1999 run only, the 230~GHz receiver was
tuned to the frequency of \COO~and the corresponding mapped area is
then smaller than in the \CO~line.

All the observations have been obtained in the frequency switching
mode \citep{lis97}, with the switch interval depending on the line
frequency. This has been chosen to be larger than the extent of the
emission (to avoid line signal overlap), but small enough to avoid
losses of emission features in the reference cycle, even if this was
not always the case for the \COO~observations, in which the spectral
range is halved with respect to the \CO.

We acquired the maps adopting a grid spacing of 50\arcsec, so that
the \CO~is slightly undersampled while the \COO~is a factor of two
undersampled. For an estimated distance $d=700$~pc, the adopted grid
spacing corresponds to a spatial scale of 0.17~pc on the cloud. The
coordinates of the (0,0) position in the map are $\alpha (2000)=
8^h48^m40^s$, $\delta (2000)=-43\degr46\arcmin12\arcsec$, and the
pointings range from $\alpha_{off}=-2000\arcsec$ to
$\alpha_{off}=+1200\arcsec$ in right ascension and from
$\delta_{off}-2000\arcsec$ to $\delta_{off}=+1950\arcsec$ in
declination. Some spatial gaps, mainly in the case of the \COO~line,
remained after our observing campaign and are described below.

The integration time at each point was generally set to
$t_{int}=10$~s, but there is a significant fraction of spectra
observed with $t_{int} = 20$~s. The typical rms noise affecting the
data (main beam temperature) is $\Delta T_{rms} \sim 0.7$~K for both
lines.

Towards many pointings, the observations were repeated on different
dates in order to check the data for consistency. In particular, the
matching between 1999 and 2003 spectra was fully satisfactory
(differences between peak temperatures are always within~10\%),
except for a raster of points located at south-east of the map, in
the $+250\arcsec\leq \alpha_{off}\leq +950\arcsec$,
$-2000\arcsec\leq \delta_{off}\leq -1050\arcsec$ region, which were
affected by temporary instrumental problems, and then partially
re-observed on subsequent dates. The reader should then be aware
that in the region $+250\arcsec\leq \alpha_{off}\leq +950\arcsec$,
$-2000\arcsec\leq \delta_{off}\leq -1700\arcsec$, these repeated
pointings are not available, so this part of the map is less
accurate.

The pointing accuracy was checked every 2-3 hours towards nearby (in
the sky) SiO masers being within $\sim 5\arcsec$.

Data reduction followed the pipeline described in \citet{mas97}:
first, spectra in antenna temperature were scaled by the main beam
efficiency factor $\eta_{mb}$, in order to express them in terms of
main beam temperature ($T_{mb}=T_A/\eta_{mb}$); then a polynomial
fit of the baseline was subtracted and a folding was performed on
the resulting spectra. Finally, spectra have been resampled to a
velocity resolution of 0.12~km~s$^{-1}$ for the \CO, and
0.06~km~s$^{-1}$ for the \COO, respectively.

In the cases of spatial superposition of repeated observations, the
corresponding spectra have been averaged (with weights depending on
the integration time and the inverse of the system temperature), in
order to obtain a better signal-to-noise ratio for the resulting
spectrum.

\section{Observational results}\label{results}

\subsection{Integrated intensity maps}
\begin{figure}
\epsscale{1.} \plotone{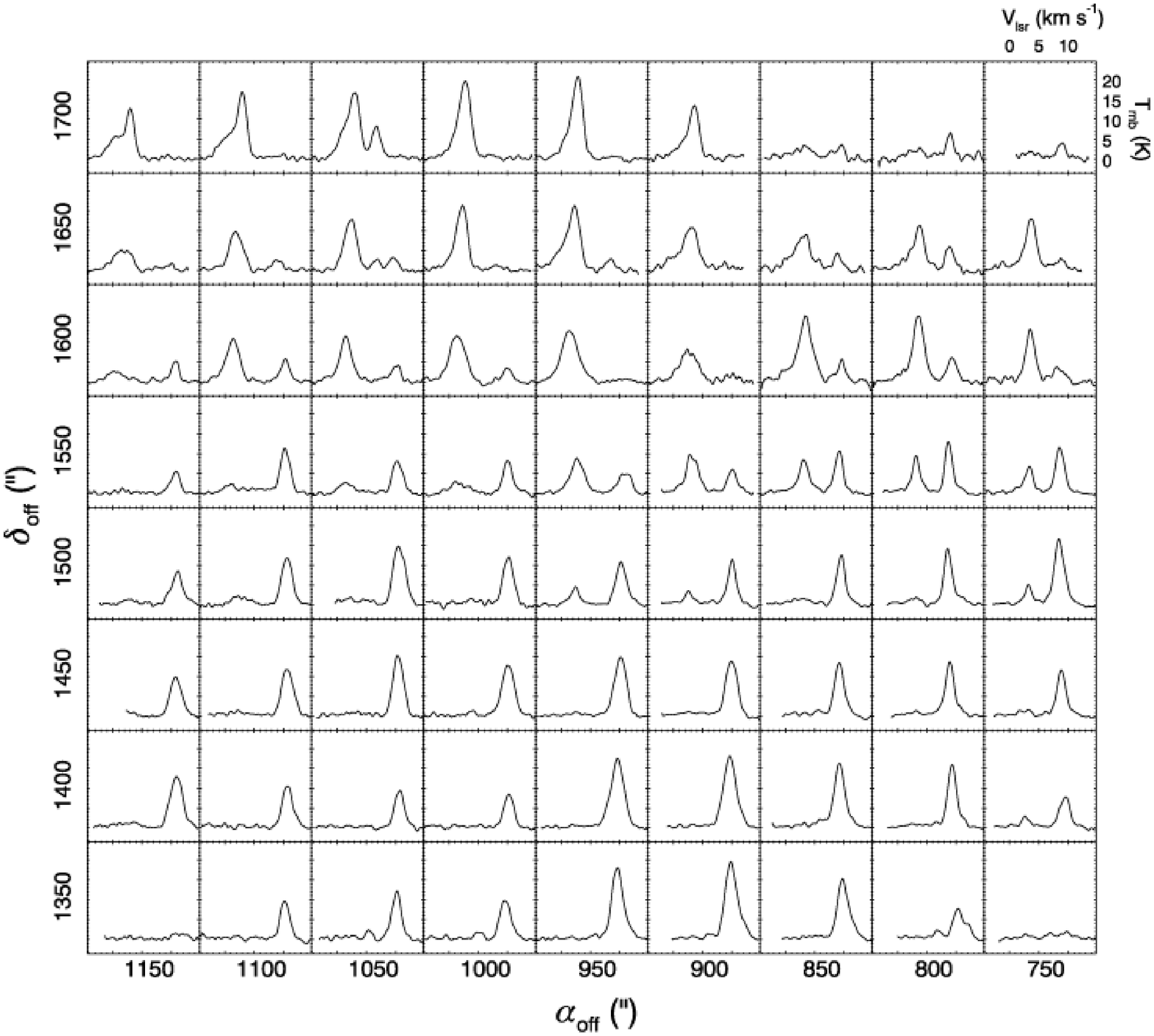} \caption{Typical line profiles. This
sample illustrates the part of the map corresponding to the offset
interval $+750\arcsec \leq \alpha_{off} \leq +1150\arcsec$,
$+1350\arcsec\leq \delta_{off} \leq +1700\arcsec$.
\label{multiplespec} }
\end{figure}

Out of the 4258 observed points, significant \CO~emission was
detected from 3392 of them, corresponding to a detection rate of
$\sim80\%$. In Figure~\ref{multiplespec}, a sample of reduced
\CO~spectra, taken towards the north-western region of our VMR-D
map, is shown. Despite this part of the grid is relatively small, it
is representative of the velocity field complexity that
characterizes this cloud. In the case of the \COO~line the points
showing significant emission at our sensitivity level are 648 out of
2393 ($\sim27\%$).

\begin{figure}
\epsscale{.5} \plotone{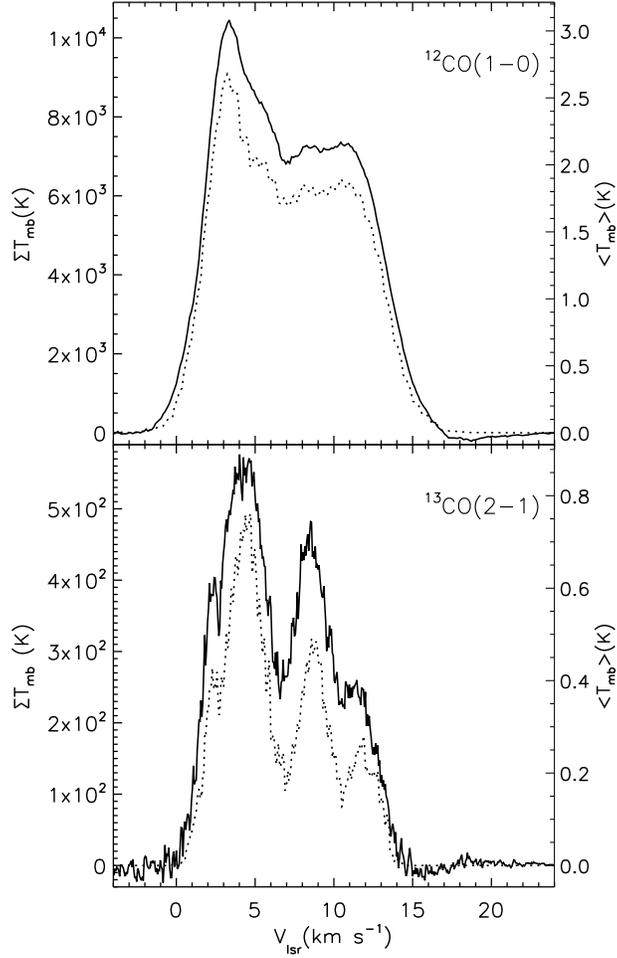} \caption{Sum of all the \CO~(upper
panel) and \COO~spectra (lower panel). The dotted line in both
panels represents the contribution of the sole volume pixels
assigned to clumps after the cloud decomposition (see
Section~\ref{clumps}). Since the solid lines can be also interpreted
as averaged spectra (by dividing by the number of points showing
significant emission), a re-scaled $y$ axis is reported on the right
to help in this reading. \label{sums}}
\end{figure}

\begin{figure}
\epsscale{1.} \plotone{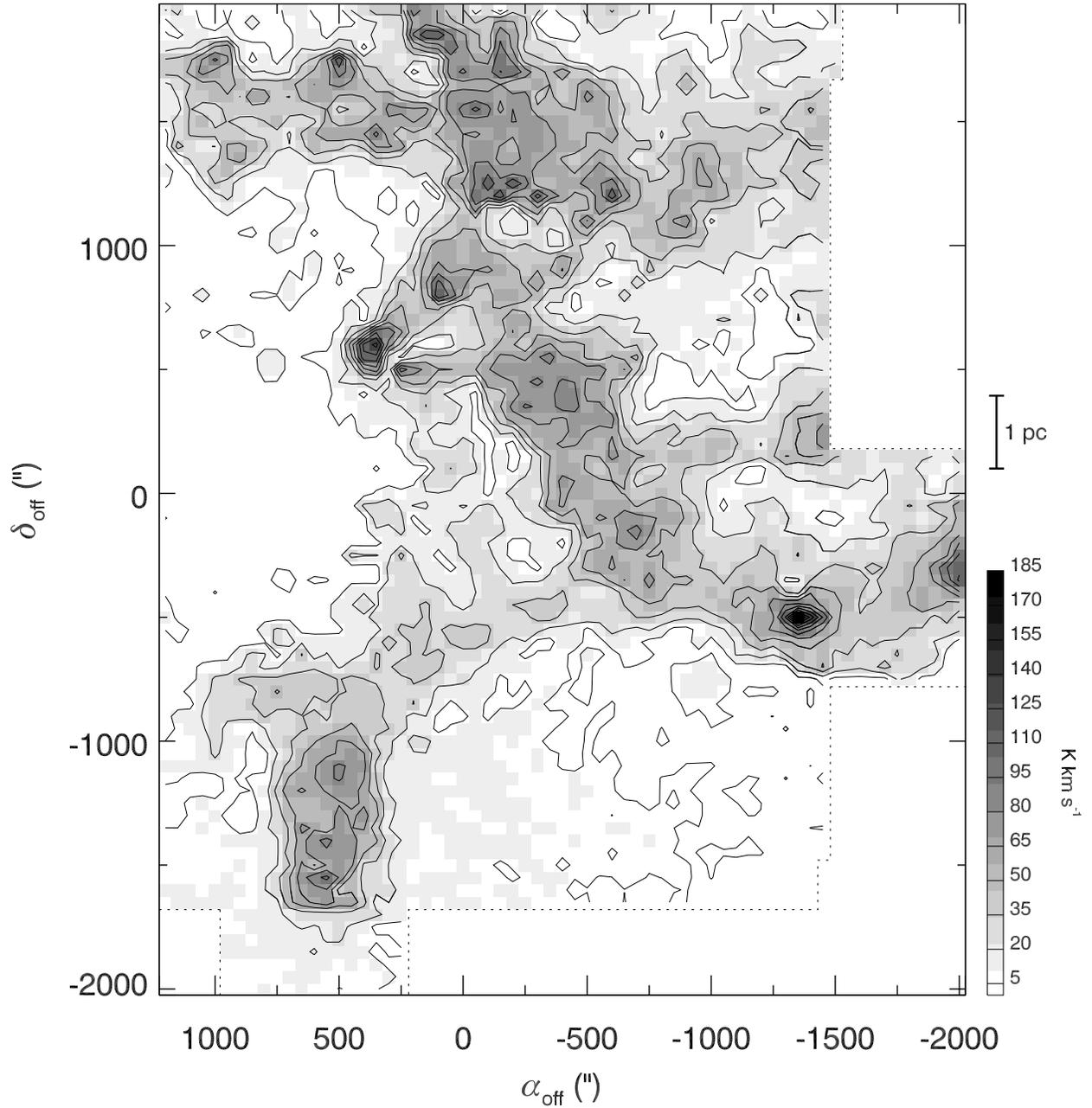} \caption{Integrated intensity map for
the \CO~emission in VMR-D region, in the range $-2\div
20$~km~s$^{-1}$. Contour levels start from 5~K~km~s$^{-1}$ and are
separated by~15~K~km~s$^{-1}$. Dashed lines delimit the observed
area. On the right, the spatial scale corresponding to the estimated
distance ($d=700$~pc) is shown. \label{map1}}
\end{figure}

\begin{figure}
\epsscale{1.} \plotone{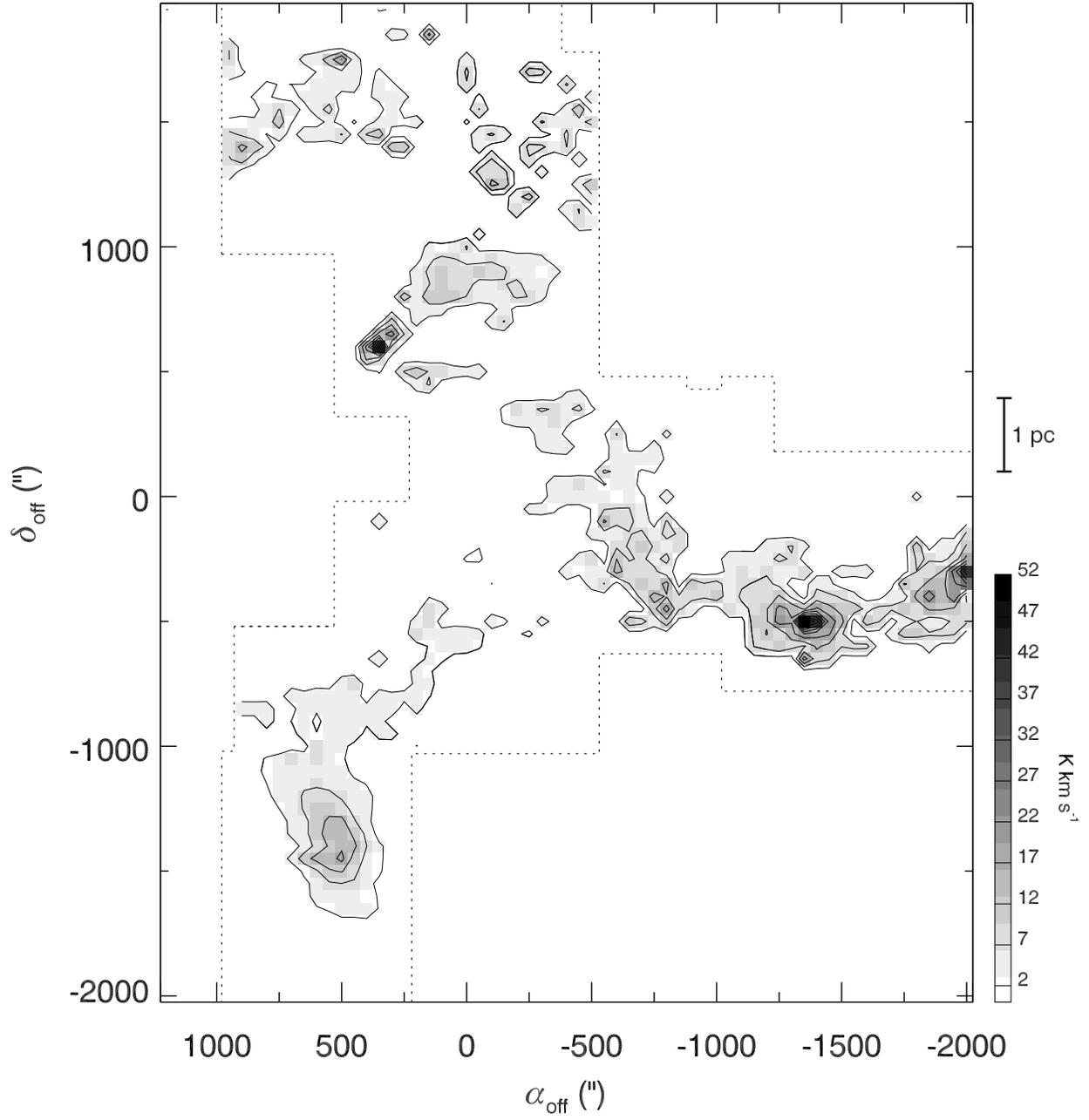} \caption{As in Figure \ref{map1}, but
for the \COO~emission. Contour levels start from 2~K~km~s$^{-1}$ and
are separated by~5~K~km~s$^{-1}$. \label{map2}}
\end{figure}

The bulk of the emission falls in the range $V_{lsr}=-2\div
20$~km~s$^{-1}$ for the \CO~line and $V_{lsr}=0\div 14$~km~s$^{-1}$
for the \COO~line, as shown in Figure~\ref{sums}, consistently with
\citetalias{wou99} and \citetalias{yam99}. The total intensity of
the \CO~and \COO~emission, i.e. $\int T_{mb}~dv$ integrated from~-2
to 20~km~s$^{-1}$, are represented in grey-scale in
Figure~\ref{map1} and Figure~\ref{map2}, respectively.

\begin{figure}
\epsscale{1.} \plotone{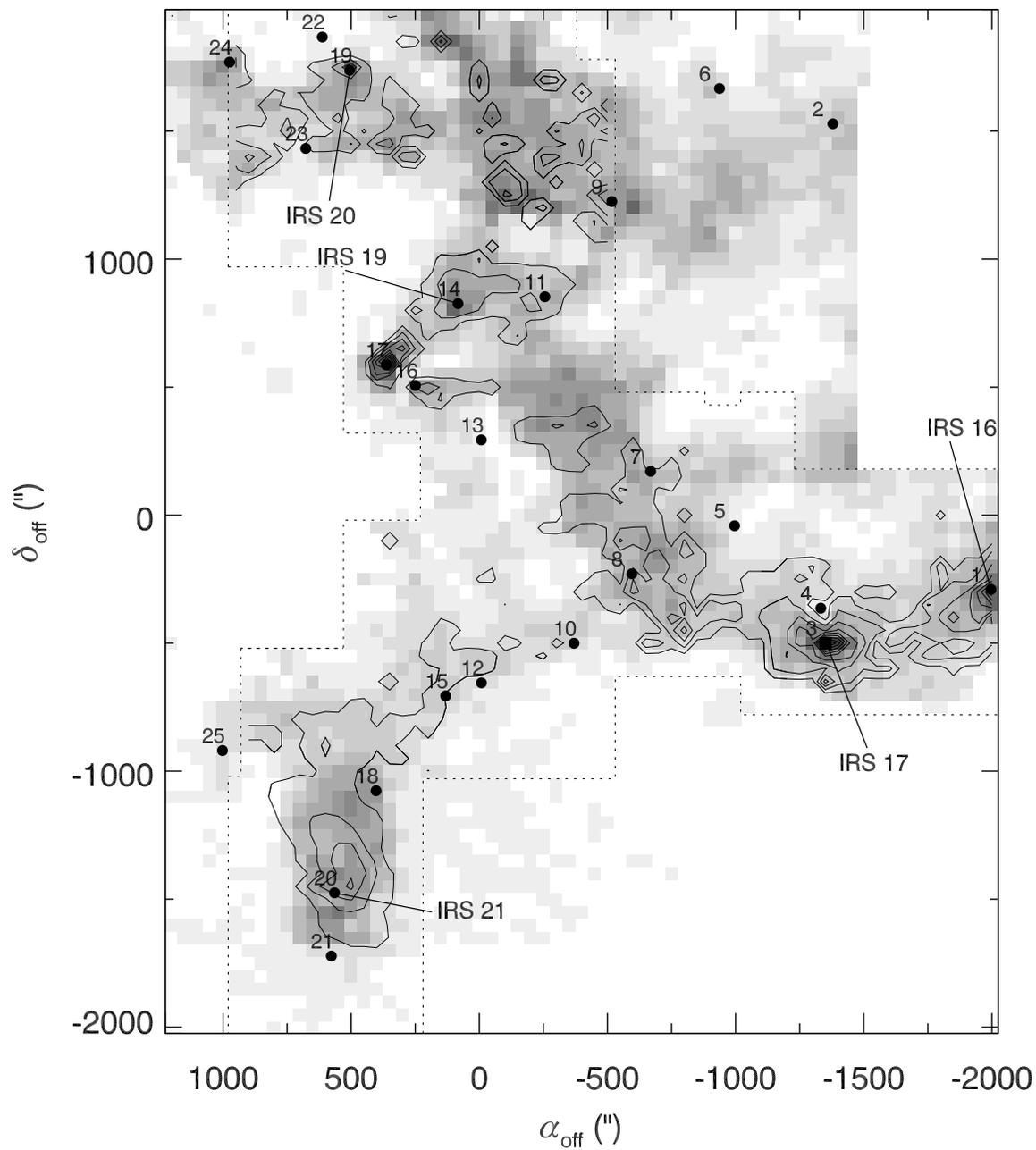} \caption{Superposition of the
\COO~contour levels (shown in Figure~\ref{map2}) on the
\CO~integrated intensity map (shown in Figure~\ref{map1}). The
positions of the ``red'' ($F_{12}<F_{25}<F_{60}$) IRAS sources are
also marked with filled circles. \label{map12}}
\end{figure}

The \CO~map in Figure~\ref{map1} evidences a complex clumpy and
filamentary structure displaying cavities of various sizes. The
analysis of the velocity components, based on the channel maps and
the vel-pos diagrams presented in Section~\ref{channels}, helps us
in separating the various emitting regions. Here we limit ourselves
to sketch the main characteristics of the cloud shape, using the
locations of the bright far-IR sources in the field as reference
points (see Figure~\ref{map12}):
\begin{itemize}
\item A large region in the northern part ($-1500\arcsec\lesssim
\alpha_{off}\lesssim +1200\arcsec$, $+1100\arcsec\lesssim
\delta_{off}\lesssim +1950\arcsec$) characterized by a strong,
distributed emission. The coincidence between the position of IRS~20
and the most intense peak in this part of the map is evident.
\item A central NE-SW region ($-2000\arcsec\lesssim \alpha_{off}\lesssim
+500\arcsec$, $-750\arcsec\lesssim \delta_{off}\lesssim
+1000\arcsec$) approximately elongated from the location of IRS~19,
associated with a bright peak of integrated emission, to that of
IRS~17 (the strongest peak in the whole map) and IRS~16, at the
western boundary. This emitting region separate three zones, east,
west, and south of it, respectively, characterized by an evident
lack of emission and determining an arc-like appearance for the gas.

\item An apparently compact structure in the south-eastern part of the map
($-250\arcsec\lesssim \alpha_{off}\lesssim +1050\arcsec$,
$-1800\arcsec\lesssim \delta_{off}\lesssim -800\arcsec$), hosting
IRS~21. This region contributes also to form the southern arc of
molecular gas.
\end{itemize}

The distribution of \COO~emission closely follows that of the more
intense \CO, as can be easily seen in Figure~\ref{map2} and
especially in Figure~\ref{map12}, in which, to highlight the
correlation between the emission of the two lines, we superimpose
the levels of the latter on the grey-scale map of the former. The
\COO~traces the densest parts of the \CO~map, and its peaks coincide
with those of the \CO, except for the main peak in the south-eastern
region. The northern region of the map is characterized by a number
of isolated peaks, some of which are also recognizable in the
\CO~map.

In Figure~\ref{map12} the positions of 25 \textit{IRAS}~PSC sources
with fluxes increasing with wavelength ($F_{12}<F_{25}<F_{60}$) are
also overplotted. Among these objects, there are also IRS~16,
IRS~17, IRS~19, IRS~20, IRS~21, characterized by the further
constraint $F_{25}>2.5$~Jy, recognized as intermediate mass YSOs by
\citet{lis92}, and associated to embedded young clusters
\citep{mas00,mas03}. Incidentally, a search for objects with
increasing fluxes in the \textit{MSX} Point Source Catalog
\citep{ega03} provided only four sources, coinciding with IRS~16,
IRS~17, IRS~19, IRS~20. The reason for this small number of selected
objects with respect to the \textit{IRAS} archive query probably
lies in the different extension of the spectral range observed.

Typically, all these sources are found in the densest parts of the
cloud, and in most cases their locations follow the arc-like
morphology of the gas. This is suggestive of a star formation
process triggered by the effects of expanding shells driven, e.g.,
by nearby massive young stars or supernova remnants. In
Section~\ref{shells} we shall explore in some detail this
possibility.

\subsection{Gas physical properties}\label{props}
The present observations have been also used to derive some physical
properties of the emitting gas. As a first approximation, the column
density of the molecular gas traced by the thermalized and optically
thick \CO~line can be determined by means of the empirical formula
\begin{equation}
N(\mathrm{H_2})=(2.3\pm 0.3)\times 10^{20} \int
T_{mb}(^{12}\mathrm{CO})\,dv~~(\mathrm{cm^{-2}})\,,
\end{equation}
which is based on a galactic average factor \citep{str88}. The
molecular mass can be derived by the relation
\begin{equation}\label{mass12}
M_{gas}(^{12}\mathrm{CO})=\mu~m_{\mathrm H}
\sum\left[d^2~\Delta\alpha~ \Delta\delta~N(\mathrm{H_2})\right]\,,
\end{equation}
\citep[see, e.g.,][]{bou97}, where $d=700$~pc is the distance of
VMR-D, $(d^2\Delta\alpha\Delta\delta)$ is the size of the emitting
area at the observed position, $m_{\mathrm H}$ is the mass of a
hydrogen atom, and $\mu$ represents the mean molecular mass.
Adopting a relative helium abundance of 25\% in mass, $\mu=2.8$ and
the derived cloud mass is $M_{gas}(^{12}\mathrm{CO})=1.5\times
10^{4}$~M$_{\sun}$. The quite large relative error on the distance
estimate, $\Delta d = 200$~pc, dominates the total uncertainty
affecting this value ($\Delta M/M \sim 50\%$), although it is
necessary to clarify that the errors implied in the assumptions
underlying this calculation probably exceed and dominate this
amount. This value represents only a small fraction ($\sim$ 2.5\%)
of the mass reported by \citetalias{yam99} for the whole complex,
i.e. an area of $\sim 100$ deg$^2$, because this is the mass of only
a small part of one out of four clouds. Note also that VMR-B
strongly contributes to the determination of the mass of the whole
complex (see Equation \ref{mass12}), being three times more distant
than the other clouds and showing a similar angular extension.

A different and more rigorous method to derive the column density
relies on the simultaneous observations of the two lines, exploiting
their different optical depth. As a first step the excitation
temperature is estimated from the peak temperature of the
\CO~emission, indicated with $T_\mathrm{R}(^{12}\mathrm{CO})$,
assuming LTE conditions and $\tau \gg 1$
\citepalias[e.g.,][]{yam99}:
\begin{equation}
T_{ex}=\frac{5.53}{\ln\{1+5.53/[T_\mathrm{R}(^{12}\mathrm{CO})+0.819]\}}\:.
\end{equation}
The peak strengths were derived by gaussian fits to the \CO~line
profiles. Often, these show very complex profiles in this region: in
these cases, we fitted only the component corresponding in velocity
to the \COO~feature, that generally shows only a single component.
Assuming that the excitation temperature is the same also for
$^{13}$CO we then determined the optical depth of the \COO~per
velocity channel by means of the relation:
\begin{equation}\label{eqtau}
\tau_v(^{13}\mathrm{CO})=-\ln\left\{1-\frac{T_v(^{13}\mathrm{CO})}{10.58}
\left(\frac{1}{\exp(10.58/T_{ex})-1}-0.02 \right)^{-1}\right\}
\end{equation}
and the corresponding column density by means of the relationship
\citep[e.g.,][]{bou97}
\begin{equation}
N(^{13}\mathrm{CO})=1.21\times
10^{14}\frac{(T_{ex}+0.88)\exp(5.29/T_{ex})}
{1-\exp(-10.58/T_{ex})}\int \tau_v (^{13}\mathrm{CO})~dv\,
\end{equation}

Finally, after N(H$_2$) is derived from $N(^{13}\mathrm{CO})$
assuming a $^{13}$CO abundance ratio of $7\times10^{5}$
\citep{dic78}, the molecular mass is calculated in the same way as
for \CO, amounting to $M_{gas}(^{13}\mathrm{CO})=1.2\times
10^{3}$~M$_{\sun}$.

Given the undersampling of \COO, a twofold bias (acting in different directions) affect the mass determination
obtained through this line. On the one hand, some emission in this line is undoubtedly lost and hence
unaccounted for in the total mass determination, but on the other hand, being \COO~less beam-diluted than \CO,
then the excitation temperature as derived from \CO~peak main beam temperature is underestimated (because of the
smaller beam filling factor), causing an overestimate in the derived column densities.

In order to compare the LTE method results with those obtained by
applying Equation~\ref{mass12}, we calculated the mass traced by
\CO~considering only those positions showing detectable
\COO~emission; we found $M^*_{gas}(^{12}\mathrm{CO})=4.8\times
10^{3}$~M$_{\sun}$.

Both the $M_{gas}(^{12}\mathrm{CO})$ and $M_{gas}(^{13}\mathrm{CO})$ masses are comparable to those obtained for
the Orion molecular cloud which possesses similar characteristics from the point of view of the star formation
activity. Of course, the comparison has to be performed taking into account both different angular size and
distance of the two clouds. Mapping in \CO~an area of 29~deg$^2$ in Orion~A and of 19~deg$^2$ in Orion~B, both
located at 560~pc \citep{gen89}, \citet{mad86} derived $M_A=1\times 10^{5}$~M$_{\sun}$ and $M_B=9\times
10^{4}$~M$_{\sun}$, respectively. Subsequently \citet{cam99} derived, from a visual extinction analysis of the
whole cloud, a total mass $M_{AB}=3\times 10^{5}$~M$_{\sun}$. Finally, a mass estimation of Orion~A through
$^{13}$CO emission, although considering the $J=1-0$ transition, has been provided by \citet{nag98}, that found
$M_{A,13}=5.4\times 10^{4}$~M$_{\sun}$. Considering then the observed areas, these values appear consistent with
those estimated for VMR-D.

\subsection{Velocity structure}\label{channels}
\begin{figure}
{\includegraphics[angle=90,scale=0.73]{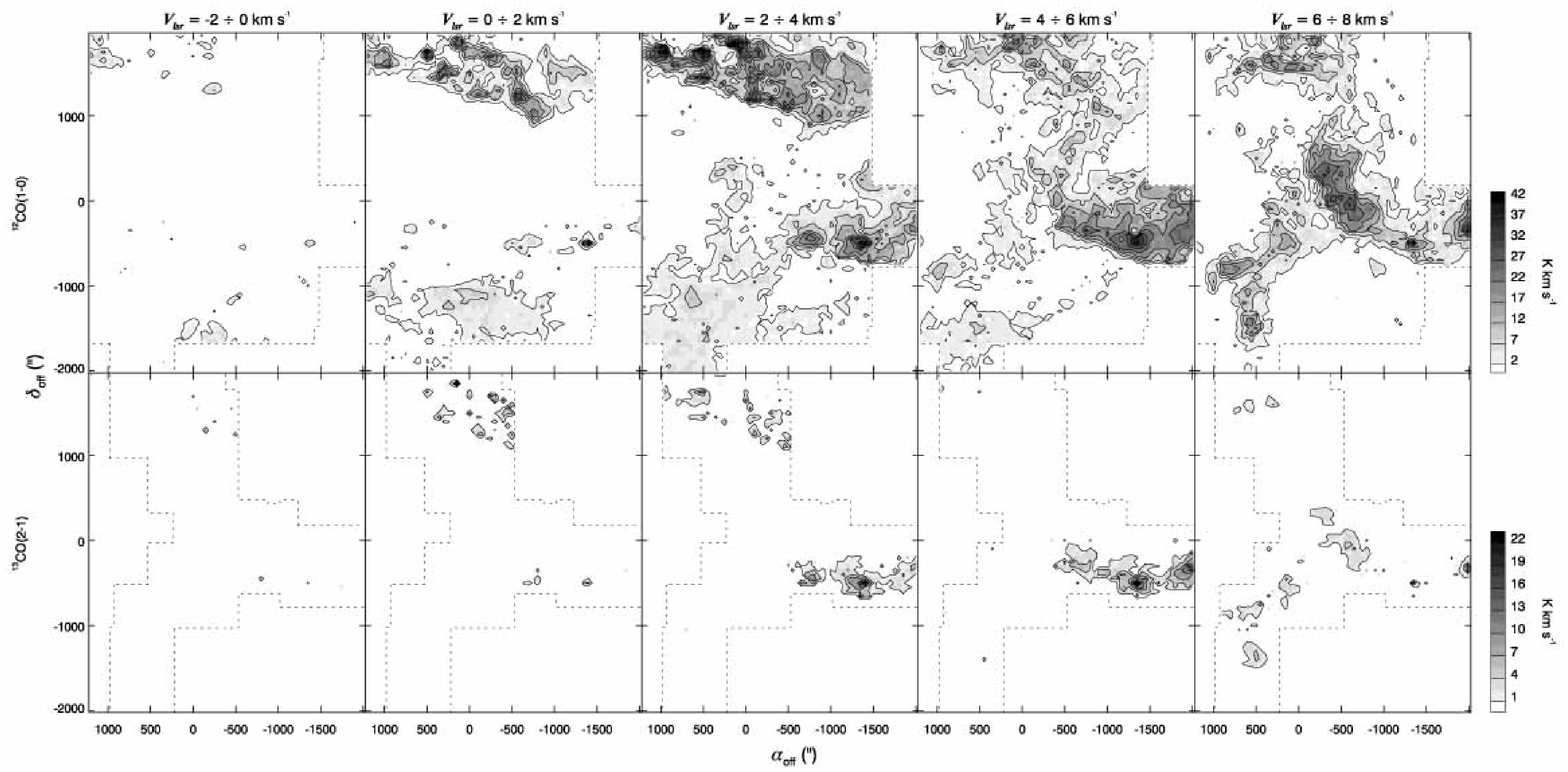}} \caption{Channel maps from -2 to 8~km~s$^{-1}$ for the \CO~(left
panels) and \COO~(right panels) emission in VMR-D region, in the velocity ranges indicated on the left border of
each box. For \CO~Contour levels start from 2~K~km~s$^{-1}$ and are separated by~5~K~km~s$^{-1}$, while for
\COO~they start from 1.5~K~km~s$^{-1}$ and are separated by~3~K~km~s$^{-1}$. In both cases, dashed lines delimit
the observed area. \label{chann1}}
\end{figure}

\begin{figure}
{\includegraphics[angle=90,scale=0.73]{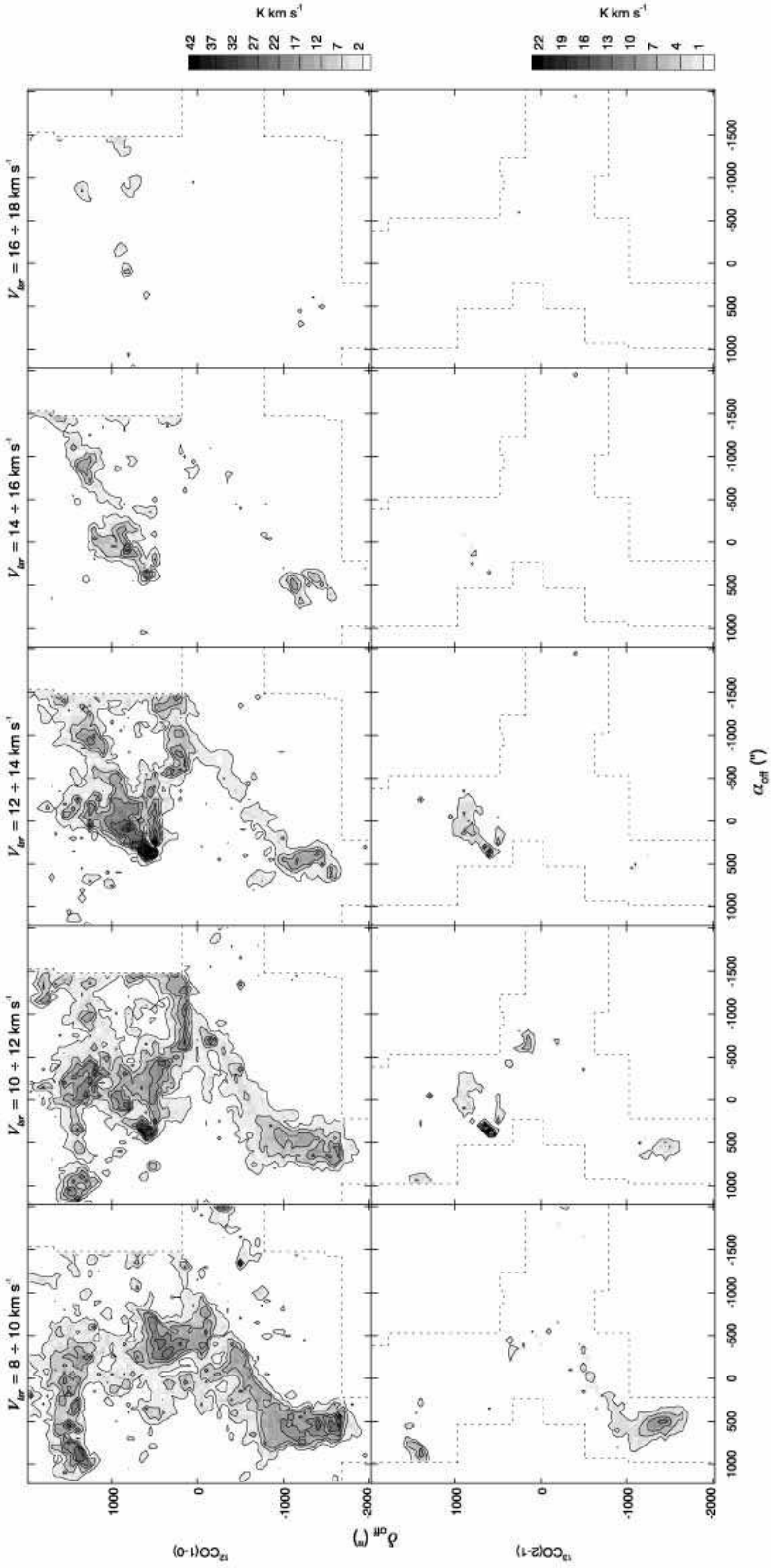}} \caption{Same as in
Figure \ref{chann1}, but from 8 to 18~km~s$^{-1}$. \label{chann2}}
\end{figure}

In Figures~\ref{chann1} and~\ref{chann2} we present the maps, for
both lines, of the integrated intensity taken in velocity intervals
of 2~km~s$^{-1}$ and in steps of the same amount. These velocity
channel maps clearly show how the emission detected in different
locations is contributed by multiple velocity components. In
Table~\ref{tabchann} the characteristic velocity ranges for the main
areas of the maps are summarized.

The \COO~channel maps, in particular, allow us to better recognize
the main velocity components of the densest regions. For example,
the emission from the northern part of the map appears clearly
separated into two components: one in the $V_{lsr}\simeq
0\div4$~km~s$^{-1}$ range, and the other in the $V_{lsr}\simeq 6\div
12$~km~s$^{-1}$ range.

Due to the large number of observed positions, as well as the geometrical regularity of the grid (in particular
for the \CO~transition), we can extract vel-pos diagrams, once an offset in $\alpha$ or $\delta$ is given and
the observed spectra along the given strip of the map are considered. Here we present 24 declination vs velocity
\CO~stripes with a fixed $\alpha_{off}$ spanning the range from $\alpha_{off}=+950\arcsec$ to
$\alpha_{off}=-1450\arcsec$ (i.e. the location of IRS~17) in steps of 100\arcsec~(Figure~\ref{velposalpha}),
that visualize the velocity field across the cloud. In principle the same could be done for the \COO~line, but
at our sensitivity level the vel-pos diagrams do not add further important information so that here we omit
them.

Many clump-like structures are visible in these diagrams, confirming
that VMR-D possesses a high degree of inhomogeneity as will be
discussed in Section~\ref{clumps}. Several cases of strong line
broadening are also detectable in correspondence to the brightest
emission peaks, as for example at $\alpha_{off}=+50\arcsec$ or
$\alpha_{off}=-1350\arcsec$, confirming the probable presence of
outflows.

The appearance of arc-shaped structures in these diagrams can be
explained either by the presence of two Galactic velocity components
corresponding to gas located at different distances or by the action
of expanding shells. In particular, this kind of structure is
detectable in the regions
$+750\arcsec\lesssim\alpha_{off}\lesssim-150\arcsec$ and
$-1800\arcsec\lesssim\delta_{off}\lesssim-500\arcsec$ (southern part
of the map), and
$+350\arcsec\lesssim\alpha_{off}\lesssim-750\arcsec$ and
$+400\arcsec\lesssim\delta_{off}\lesssim+1800\arcsec$ (northern part
of the map).

The first hypothesis implies extremely different kinematical
distances ($\sim 1$~kpc) between the two emitting components, an
occurrence indicating the presence of uncorrelated clouds along the
same line of sight. The other possibility of the shell expansion is
intuitively suggested by the arc-like shapes (in
Figure~\ref{velposalpha} we draw ellipses to show four possible
cases), even if this can be also justified by simply invoking the
internal motions of the cloud, a scenario that seems more
appropriate as will be discussed in next section.

\subsection{Expanding shells}\label{shells}
\begin{figure}
\epsscale{.9} \plotone{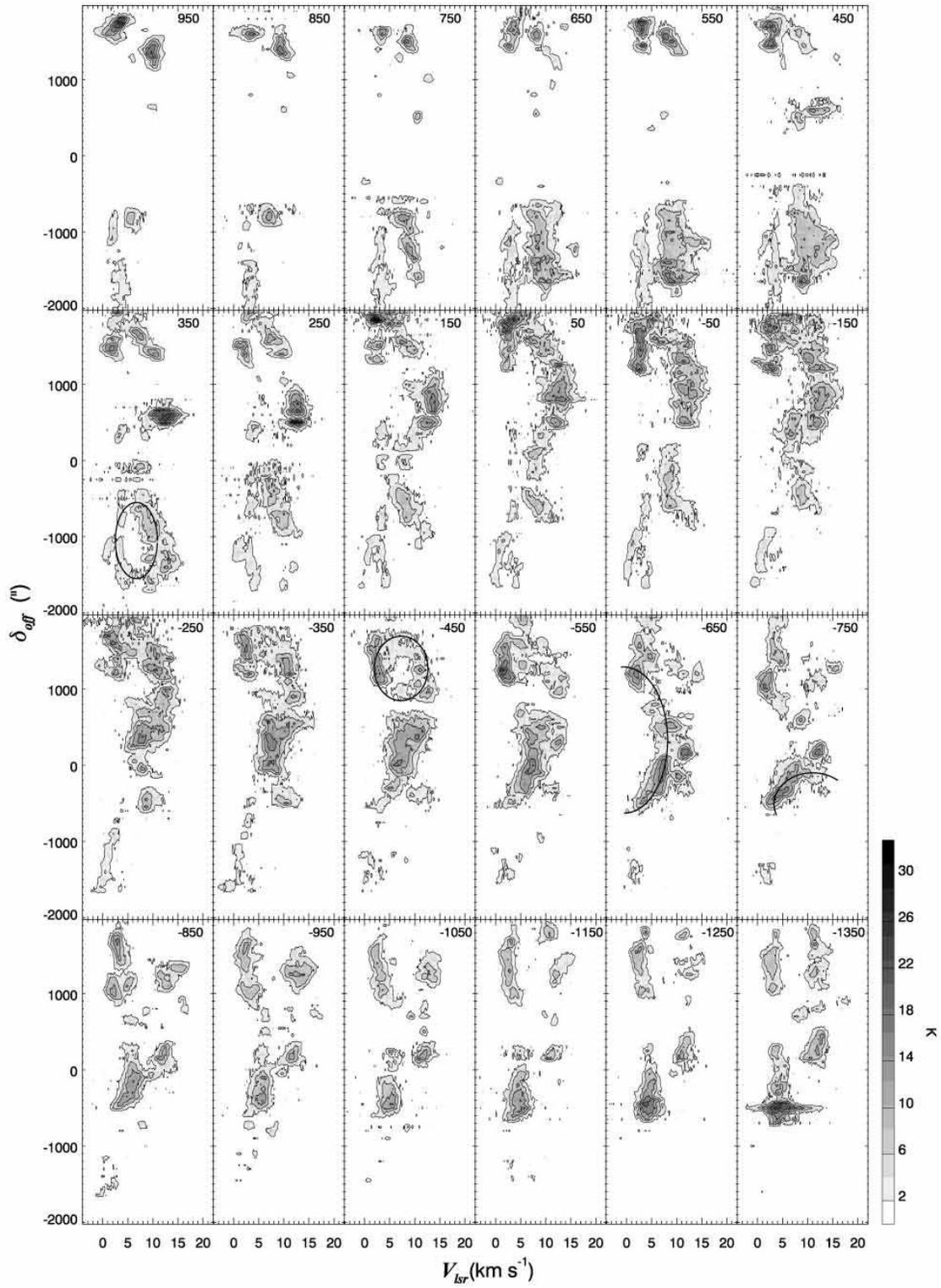} \caption{Vel-$\delta_{off}$ diagrams
of the \CO~emission in VMR-D, for fixed values of $\alpha_{off}$,
indicated in each panel. Contour levels start from $T_{mb}=2$~K and
are separated by~4~K. Four arc-like structures have been recognized
and marked with a solid arc of ellipse, each one in the diagram in
which it is best visible. \label{velposalpha}}
\end{figure}

The presence of arc-like features in both $\alpha-\delta$
(Figures~\ref{map1}, \ref{map2}, \ref{map12}) and vel-$\delta$ maps
(Figure~\ref{velposalpha}) of VMR-D region could be interpreted as a
signature of possible expanding shells, a phenomenon that is often
related with star forming activity.

\begin{figure}
\epsscale{.85} \plotone{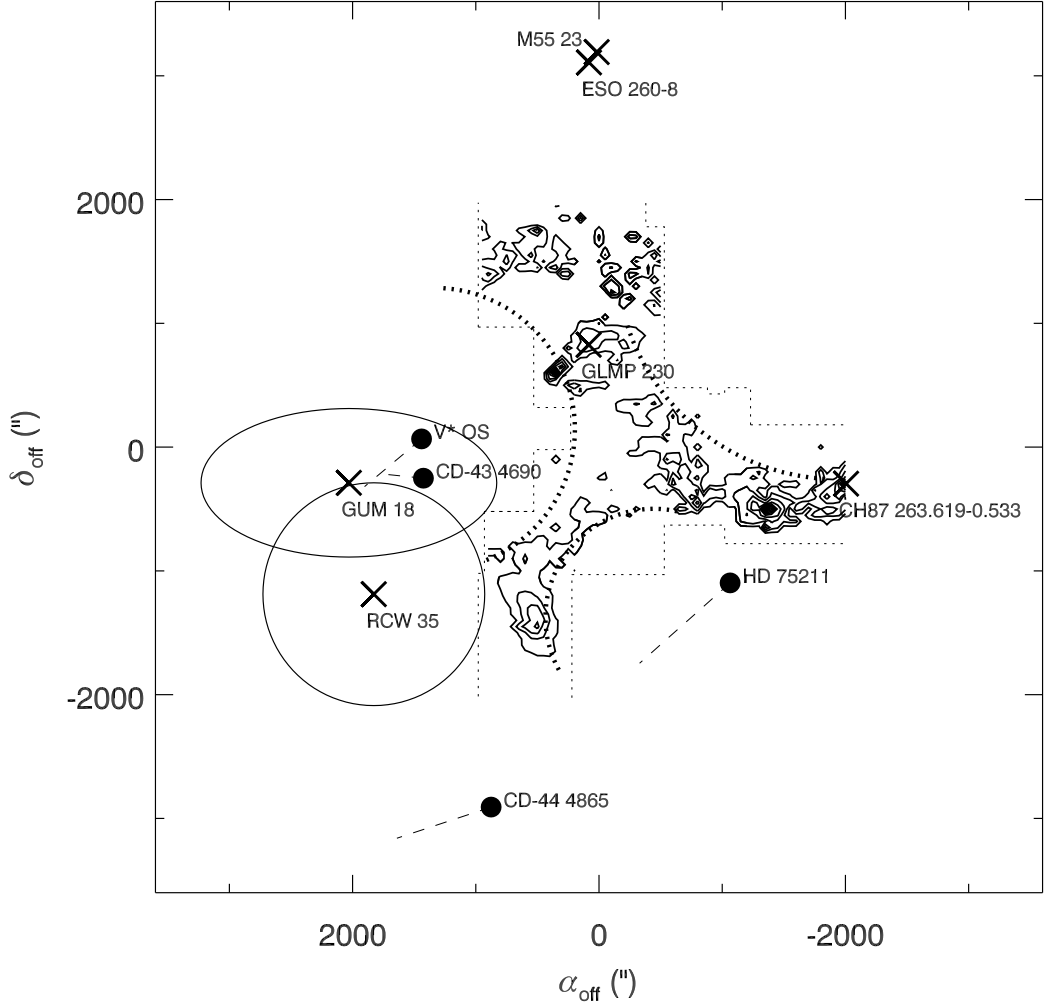} \caption{A $2\degr\times2\degr$
field centered on the (0,0) position of the map is shown, with the
contours of the \COO~emission plotted as in Figure~\ref{map2}. The
bold dotted circles roughly indicate three arc-like emitting
regions, while the thin dashed lines mark the border of the observed
zone. Crosses indicate the center of the \textsc{Hii} regions; for
two of them with significant spatial extent, Gum~18 and RCW~35, an
ellipse is also drawn whose axes length is taken from literature.
The remaining \textsc{Hii} regions are compact objects and cannot be
responsible for the large-scale morphology. The locations of OB-type
stars (filled circles), with their proper motion track extrapolated
for the last $10^5$~yr, are also marked. \label{obstars}}
\end{figure}

To explore this possibility we considered the available information
on the distribution of the known OB stars and \textsc{Hii} regions
in the VMR-D field, that are shown in Figure~\ref{obstars} along
with the spatial distribution of the gas. Three arc-like molecular
structures are marked with dashed lines, but only in the case of the
eastern arc we could guess an association with clear driving sources
candidates. A possible projective correspondence can be established
between the Gum~18 \textsc{Hii} region \citep{gum55} and this arc,
that could be tentatively related to it, but not for RCW~35
\citep{rog60}. However, the centroids of the two structures do not
coincide and the distance of Gum~18 is unknown. In addition, the OB
stars V* OS Vel and CD-43 4690 are not far from the apparent
centroid of the molecular arc, but their distances \citep[$1.7\pm
0.3$~kpc in both cases,][]{rus03} are inconsistent with that
estimated for VMR-D, so that we conclude that they cannot be
responsible of the arc-like morphology of the gas. Similar
considerations apply in the case of the possible association between
the southern arc and the location of HD~75211
\citep[$d=1.2$~kpc,][]{sav85}.

\begin{figure}
\epsscale{.55} \plotone{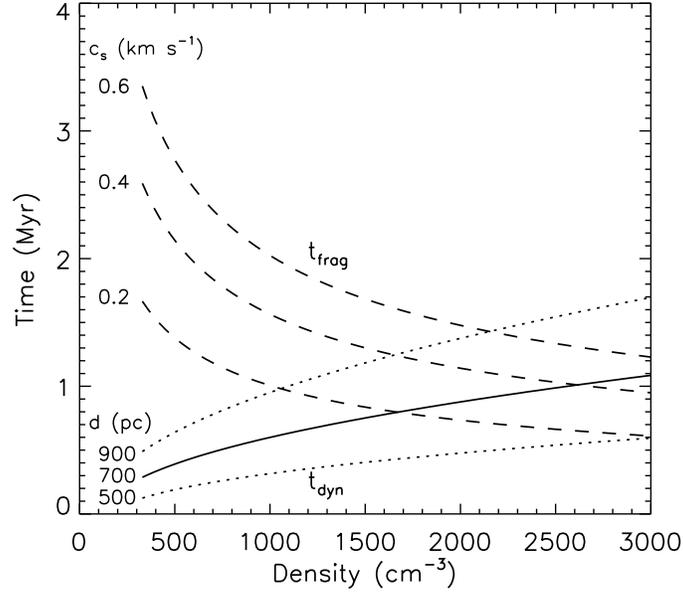} \caption{Plot of the dynamical age
$t_{dyn}$ (solid line for $d=700$~pc, and dotted lines for
$d=500$~pc and $d=900$~pc, respectively) for a \textsc{Hii} region
and of the time at which the fragmentation of the driven shell
starts (for different values of the sound velocity $c_s$ and a
radius of $\sim 3.4$~pc) as a function of the density $n_0$.
\label{tdyn}}
\end{figure}

By simple models, describing the expansion of a \textsc{Hii} region
and the fragmentation of the shocked dense layer surrounding it, it
is possible to estimate the timescales involved in the hypothesis
that the shell driving sources are \textsc{Hii} regions. The
relation between the radius $R$ of the shell, expanding in a
homogeneous and infinite medium, and its lifetime is
\begin{equation}
t_{dyn}=0.559\times \frac{R_0}{c_{\mbox{\textsc{ii}}}}\left[
\left(\frac{R}{R_0}\right)^{7/4}-1\right]~\mathrm{Myr}
\end{equation}
\citep{spi78}, where $R_0$ is the Str\"omgren sphere radius (in pc),
and $c_{\mbox{\textsc{ii}}}$ the isothermal sound speed in the
ionized region (in km~s$^{-1}$). Being the molecular arc $\sim
1000\arcsec$ in radius, the corresponding spatial scale is $R \simeq
3.4 \pm 1.0$~pc at the estimated distance of VMR-D, where the
uncertainty is related to the distance. For a star emitting
$10^{49}$ ionizing photons per second, we find $R_0=0.677~
n_3^{-2/3}$~pc, where $n_3=n_0/(1000~\mathrm{cm}^{-3})$ and $n_0$ is
the ambient density. Assuming the sound speed
$c_{\mbox{\textsc{ii}}}=10$~km~s$^{-1}$ and varying the density in
the range $n_0= 300 \div 3000$~cm$^{-3}$, we find $t_{dyn}\sim
0.2\div 1.1$~Myr. The scaling of the age with the density is shown
in Figure~\ref{tdyn} for different estimates of the distance.

Considering now the \citet{wit94} model describing the fragmentation
of the molecular gas under the action of an expanding \textsc{Hii}
shell, it is also possible to determine the time elapsed before the
onset of this phenomenon. For the considered star this model gives
\begin{equation}
t_{frag}=1.56~c_{.2}^{7/11}~n_3^{-5/11}~\mathrm{Myr}
\end{equation}
where $c_{.2}=c_s/(0.2~\mathrm{km~s}^{-1})$, and $c_s$ is the sound
speed in the molecular gas, an important parameter of the model we
chose in the range $c_s = 0.2 \div 0.6$~km~s$^{-1}$.

Because both fragmentation and star
formation appear to be already in progress, the relation between
these two times have to be $t_{dyn}
> t_{frag},$ and this happens when $t_{dyn} \gtrsim 8\times 10^5$~yr and
$n_0 \gtrsim 1600$~cm$^{-3}$, if we consider the preferred values
$d=700$~pc and $c_s=0.2$~km~s$^{-1}$. The effect of a different
distance is also shown in Figure~\ref{tdyn}.

We note that these timescales are consistent with the age of the
cluster formation activity in this region, estimated to be within
$1\div 6 $~Myr by \citet{mas06a}. This makes possible that both arcs
and the star formation activity in them are caused by the expansion
of \textsc{Hii} regions. However, since no clear excitation source
candidates have been found in the vicinity of the arc centers, we
cannot exclude different possible origins for the shells, such as
for example stellar winds from a previous generation of young stars.

Despite another possibility is suggested by the well-known Vela SNR,
apparently interacting with the whole VMR on larger scales
\citep[see Figure~1 in][]{mor01}, a correlation with the southern
arc in VMR-D is far from clear. Furthermore, considering the
possible connection between this molecular arc and its star
formation activity, there is a clear inconsistency between the age
of the YSOs and that estimated for the Vela SNR
\citep[$t_{\mathrm{SNR}}\simeq 10^4$~yr, see, e.g.,][]{mor01}. Two
further remnants, SNR~266.3-01.2 and SNR~260.4-03.4, are too far for
being responsible for the peculiar morphology of the investigated
region.

In this respect, the hypothesis that the fluidodynamic evolution of
the gas can be responsible of the observed filamentary condensations
can not be rejected.

As mentioned above, also the analysis of the cloud velocity field
can provide useful information about the presence and the physical
properties of possible expanding shells. For example, the
shell-shaped structure highlighted in Figure~\ref{velposalpha} at
$\alpha_{off}=+350\arcsec$ can be associated, due to its spatial
location, to the southern arc identified in the $\alpha-\delta$ map.
It is characterized by a peak separation $\Delta V_{lsr}\simeq
7$~km~s$^{-1}$, corresponding to an expansion velocity
$V_{exp}\simeq 3.5$~km~s$^{-1}$. Adopting a spherical symmetry, we
estimate a mean radius $R_{sh}\simeq 5.5$~pc and a corresponding
dynamical age $t_{dyn} \simeq 1.5$~Myr, a value consistent with the
possible timescales calculated above for an expanding \textsc{Hii}
region. A rough estimate of the kinetic energy can be also obtained
by adopting the assumption inspiring the Equation~\ref{mass12} and
considering the expansion velocity of each spectral component. In
this way we find $E_{kin}\simeq 2.3 \times 10^{47}$~erg, a value
comparable, for example, with the kinetic energy of a \textsc{Hii}
region possessing the same radius and age \citep{spi78}. The issue
of the identification of a driving source for this shell remains
open, as already discussed when we examined the corresponding arc
observed in the $\alpha-\delta$ map.

Another shell-like structure is well recognizable in the northern
part of the vel-pos diagrams, best visible at
$\alpha_{off}=-450\arcsec$. With considerations similar to the
previous case, we find $V_{exp}\simeq 4.5$~km~s$^{-1}$,
$R_{sh}\simeq 7$~pc, $t_{dyn} \simeq 1.5$~Myr, and $E_{kin}\simeq
5.4 \times 10^{47}$~erg. Also in this case the identification of a
possible driving source is uncertain: some \textit{IRAS} red
sources, namely those identified as 9, 11, and 14 in
Table~\ref{iraspsc}, are located in the neighborhoods of the shell,
but these objects might be presumably the product of a star forming
activity triggered by an expanding shell, rather than the cause of
it. In conclusion, the ages derived for these two candidate shells
are similar, and support the hypothesis of a synchronous
intermediate-mass star formation in this cloud, originated by a
strong compression of the gas due to the associated shocks. For this
process we suggest an upper limit in timescale of $t \simeq
1.5$~Myr.

In Figure~\ref{velposalpha} two further arc-shaped structures have
been marked with solid lines, around $\alpha_{off}=-650\arcsec$ and
$\alpha_{off}=-750\arcsec$, respectively. Although less evident than
the cases considered above, these shapes support the hypothesis of a
coupling between the arcs recognized in the integrated intensity
maps (in these cases, the western and the southern one,
respectively) and those present in the vel-pos diagrams.

\subsection{Search for outflows}
Since several positions in the map show a clear broadening of the
line profile (see, e.g., the diagram at $\alpha=-1350\arcsec$ of
Figure~\ref{velposalpha}) we studied these positions in more detail
because this is a typical signature of the presence of energetic
mass outflows \citep[see, e.g.,][]{bac96}.

\begin{figure}
\epsscale{1.} \plotone{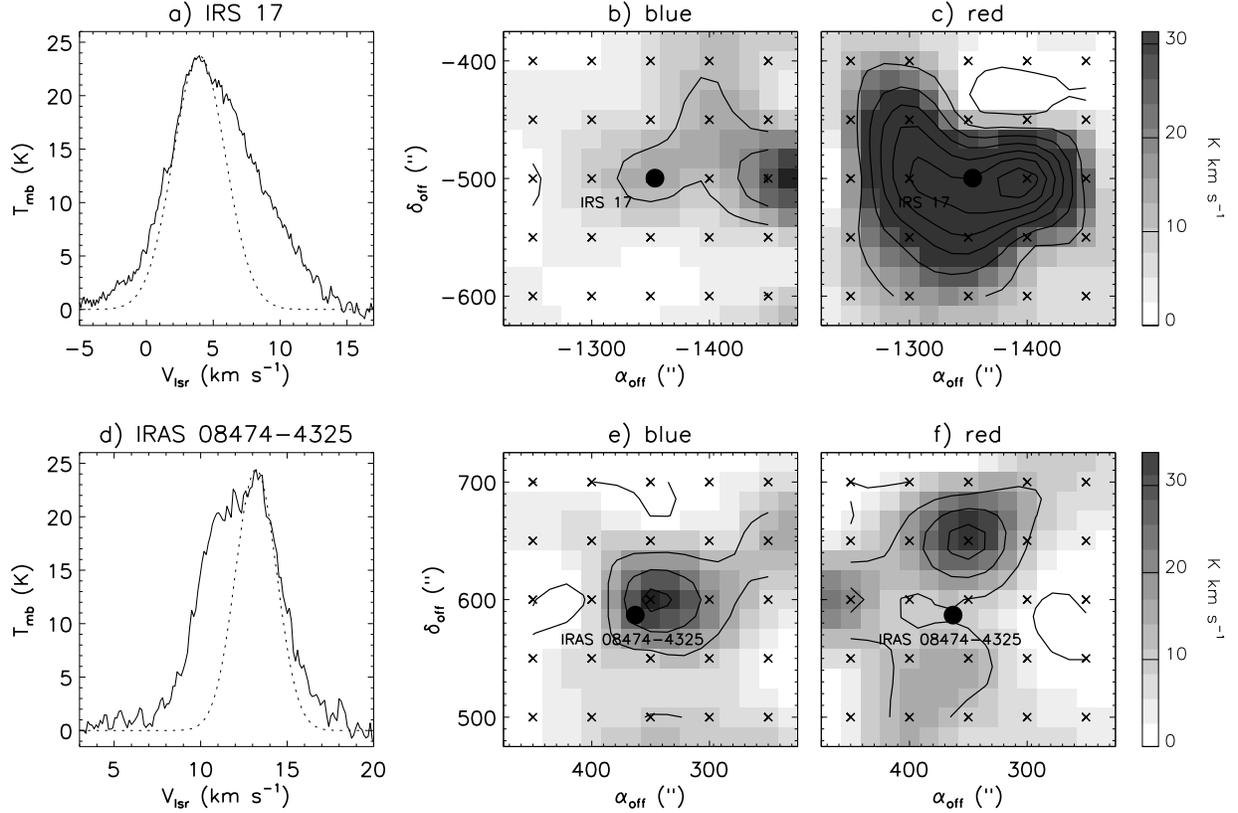} \caption{Panel $a$: \CO~spectrum
corresponding to the closest pointing to the IRS~17 location, at map
offset $(\alpha_{off}=-1350\arcsec, \delta_{off}-500\arcsec)$; the
dashed line is the gaussian determined as in \citetalias{wou99}.
Panel $b$ and $c$: interpolated greyscale maps for blue and red
outflow components, respectively. Greyscale and contour levels are
the same in both maps; the levels are in steps of~5~K~km~s$^{-1}$.
Observed positions (crosses) and IRS~17 location (filled circle) are
also shown. The blue wing peak assigned to the outflow is coincident
with the central position in panel $b$, while the brighter feature
at $100\arcsec$ on the right is considered unrelated. Note that the
red wing is probably contaminated by the contribution of a second
component to the line profile. Panels $d$, $e$, $f$: the same as
panels $a$, $b$, $c$, but for IRAS~08474-4325 (even here the blue
wing is probably contaminated). \label{outf4235}}
\end{figure}

To this aim we considered the locations of the 25 \textit{IRAS}
sources shown in Figure~\ref{map12} and quoted in
Table~\ref{iraspsc}, carrying out a systematic search for outflows
in the \CO~line toward these objects. The detection method is the
same as adopted by \citetalias{wou99}: a gaussian is fitted to the
upper part of a spectral feature, centered on the line peak, and at
the same intensity; the smaller of the two intervals between the
peak velocity and the velocities at half maximum is taken as the
HWHM of the fitting gaussian. When the profile wings significantly
exceed the gaussian wings, the presence of an outflow is inferred
and the blue and red components are obtained by integrating the
difference $T_{mb}-T_{fit}$ between the half maximum of the gaussian
and the velocity where the line wing fades into the noise. In
Figure~\ref{outf4235}, panels $a$ and $d$, we show, as an example,
the \CO~spectrum of the pointings closest to the IRS~17 and
IRAS~08474-4325 (identified with 3 and 17 in our list, respectively)
locations, with the fitting gaussian determined as described above.
In panels $b$-$c$, $e$-$f$ the intensities of the blue and red
components, in the neighborhoods of these two sources are
represented in greyscale.

In seven cases the presence of one or more additional spectral
components blended with the main feature made impracticable a clean
detection of any outflow that might be present. This difficulty
restricts our sample, even if it remains still meaningful in the
perspective of providing good target candidates for future APEX and
ALMA observations.

In 13 out of the remaining 18 cases, we detected a possible outflow, as summarized in Table~\ref{iraspsc}:
column~9 contains a flag indicating the result of the detection, column~11 the width of the velocity range
subtended by the wings, and column~12 the offset of the wing maximum emission from the map pixel in which the
source falls. Note that in two cases only one wing is detected and that when the detection flag ``Y'' is
followed by an asterisk, a gaussian fit of a secondary (but well-separated) component has been subtracted to
correctly estimate the wing intensities.

Unfortunately, the complexity of the \CO~spectra in this region
makes sometime difficult an accurate detection of the wings by
simply applying the \citetalias{wou99} method. The example of IRS~17
in Figure~\ref{outf4235}, panel~$a$, is illustrative: the
temperature peak does not stand in the center of the feature, so
that a much larger amount of emission is recognized in the red wing.
Instead, the blue wing seems better suited to be interpreted as an
outflow. Vice versa in the case of IRAS~08474-4325: the two cases
discussed have been shown in Figure~\ref{outf4235} just to highlight
the differences between a reliable detection of outflows (blue wing
of IRS~17 and red wing of IRAS~08474-4325), and more controversial
cases (red wing of IRS~17 and blue wing of IRAS~08474-4325). The
latter are indicated in Table~\ref{iraspsc} with a dagger mark.

Because our work was not focused on the study of the outflows, we
remark that the grid spacing of 50\arcsec~is quite coarse for a
detailed description of the outflow morphology, and a larger
integration time would have been required for each spectrum to
better disentangle the wing tails from the noise. These problems are
evident in comparing our results with those obtained by
\citetalias{wou99} for the three objects in common with our sample
(associated with IRS~17, IRS~19, IRS~20 respectively): in our case
the wing width is significantly smaller than found by
\citetalias{wou99}, ranging from 32\% (blue wing of IRS~20) of their
values to 71\% (red wing of IRS~17).

Furthermore, considering that \COO~observations are too noisy for a
similar analysis, it is almost impossible to exploit them for a
quantitative study of the outflow physical properties. In any case
our analysis significantly improves the outflow statistics in the
observed region, from three to 13 objects, providing a larger sample
of targets for future higher resolution observations.

\section{Cloud decomposition} \label{clumps}
To characterize the structure of VMR-D and the distribution of its
velocity components in a quantitative way we applied cloud
decomposition techniques to our data, a tool typically adopted to
study possible hierarchies in the molecular cloud structure. In
particular, we adopted the 3D version of the CLUMPFIND (hereinafter
CF) algorithm \citep{wil94}, running on a $\alpha$-$\delta$-$v$ cube
containing all the reduced spectra for each line. Important
adjustable parameters in the CF scanning procedure are the radiation
temperature threshold $T_{min}$ and the level increment $\Delta T$,
the first one being generally set slightly above the noise
fluctuations. This is an important point that requires accurate
tests because larger and smaller thresholds involve the loss of
faint peaks and the inclusion of noise peaks, respectively. The
second parameter $\Delta T$ controls the separation of close peaks,
as clearly discussed in \citet{bru03}.

A large number of tests have been run on the \COO~data cube, which
is characterized by a smaller number of features and a reduced
diffuse emission, two characteristics that allow a better control
and verification of the reliability of the clump assignments. To
obtain a convenient reduction of the noise effect and to ensure some
stability to the results, the spectra were preliminarily resampled
in velocity by a factor of 5.

\begin{figure}
\epsscale{1.} \plotone{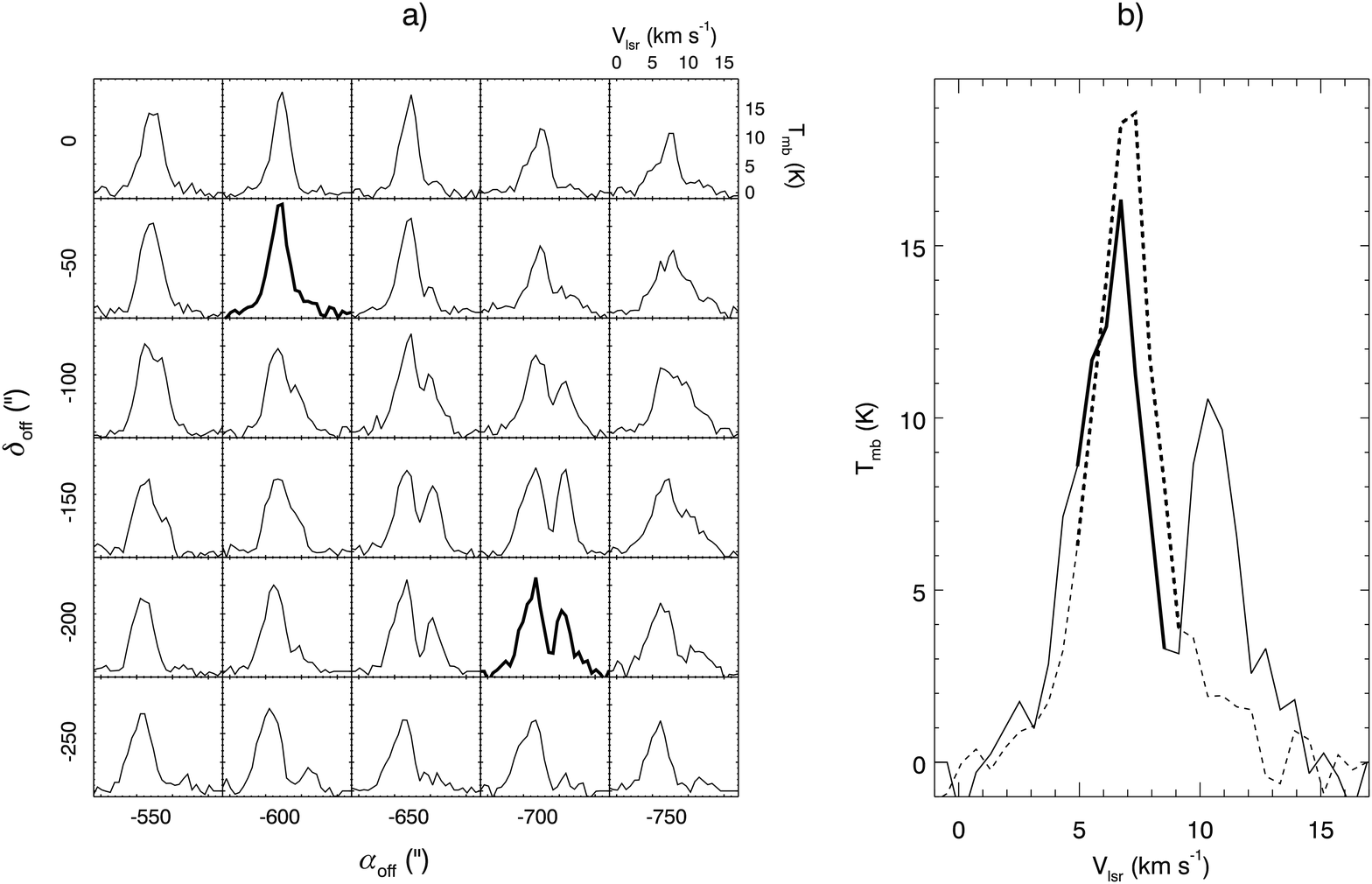} \caption{Example of bad clump
detection performed by CF. Panel~$a$: a grid of \CO~spectra
(resampled in resolution from 0.12 to 0.6~km~s$^{-1}$) around two
positions (marked with a thicker line) recognized by CF as centroids
of different clumps, with almost coincident peak velocity
($V_{lsr}\simeq 7$~km~s$^{-1}$). Because these components satisfy
our merging criteria (see text), we conservatively chose to consider
them as a single clump. Note that a second component at
$V_{lsr}\simeq 10$~km~s$^{-1}$ is clearly resolved in some spectra
and is correctly assigned by CF to another clump. Panel~$b$: the two
thicker spectra shown in panel~$a$ are superimposed and plotted with
solid and dashed line, respectively. The velocity channels
erroneously assigned by CF to two different clumps are highlighted
(bold line). \label{cferr}}
\end{figure}

Exploring the input parameter space, we noted that the choice of the
parameter set can produce not only systematic (as those described
above), but also random effects. These are particularly evident, for
example, in large clumps corresponding to a single bright line, that
are sometimes decomposed in two or more clumps, an effect that is
illustrated in Figure~\ref{cferr}. We verified that, adopting a
different set of parameters, the clump could appear as single, the
price being that some other cases, previously recognized as single,
now appear as separated. This experience suggests that the tuning of
the parameter values is critical with respect to the number of
detections. This kind of problems in cloud decomposition with CF
have been extensively discussed in \citet{bru03}, \citet{ros05}, and
\citet{ros06}.

In this work, the relatively low number of detected clumps suggested
us a very simple approach based on the careful inspection of a large
number of CF outputs. First of all, we ran CF adopting a parameter
set ensuring a good detection of the features, i.e. a $T_{min}$
value as low as possible, although above the noise limit, and a
similar value also for $\Delta T$, to minimize the effect of noise
fluctuations affecting the spectra. Both values have been set to
2~K, corresponding to $\sim 3\sigma$ in our map. In this way, the
cataloged emission accounts for the 66\% of the total emission in
the \COO~map, as shown in Figure~\ref{sums}.

The obtained clump list has then been analyzed looking for those
objects that, despite they are quite close in the
$\alpha$-$\delta$-$v$ space, clearly appear as spuriously splitted
by CF. We merged together these clumps to obtain a final clump
distribution as free as possible from the most apparent
misinterpretations of CF. In this way, we do not claim to determine
the ``true'' clump distribution in the cloud, but we believe that
the census resulting from our clump merging method is more reliable
than the simple CF output.

Our experience in analyzing the clump merging suggested a set of
constraints that have been included in an automatic merging
algorithm. In fact, we merged only clump pairs whose centroids
satisfy all the following criteria: i) both distances in $\alpha$
and $\delta$ are less or equal to 3 grid elements (corresponding to
a spatial scale of $\sim0.5$~pc); ii) the difference between the
central velocities is less than or equal to 3 channels (i.e. $\Delta
v \leq 0.9$ km s$^{-1}$, after resampling); iii) the
$\alpha$-$\delta$-$v$ space total distance
$d=\sqrt{\Delta\alpha^2+\Delta\delta^2+\Delta v^2}$ is less than or
equal to 5 volume pixels. In this context the pairs showing a
spatial discontinuity in their emission have been taken as
separated.

After this procedure the number of detected \COO~clumps decreased
from 67 to 49 clearly modifying the resulting mass spectrum, with
the total detected mass remaining obviously unchanged.

The list of these clumps, along with their coordinates, velocity
centroids, radii, masses, corresponding virial masses and optical
depths is presented in Table~\ref{clumptable}.

The clump masses have been evaluated assuming the LTE condition, as
we did in Section~\ref{props} for the whole cloud, so that the total
mass assigned to these clumps amounts to 845~M$_{\sun}$,
corresponding to the 77\% of the whole $M_{gas}(^{13}\mathrm{CO})$
calculated.

The virial masses in column~8 are obtained from the observed radius $R$ and velocity dispersion $\Delta V$ under
the hypothesis of constant density \citep{mac88}
\begin{equation}
M_{vir}=210\left(\frac{R}{\mathrm{pc}}\right) \left(\frac{\Delta
V}{\mathrm{km~s}^{-1}}\right)^2\:.
\end{equation}
For density radial profiles as $\rho\propto r^{-1}$ and $\rho\propto
r^{-2}$ the multiplicative constant decreases to 190 and 126,
respectively.

The optical depth is calculated along the line of sight of the clump centroid using Equation~\ref{eqtau}. We
list in column 9 the optical depth at peak velocity, and in column 10 the integral of this observable over the
velocity channels assigned to the considered clump. Note that for the clump VMRD43 the maximum and the
integrated optical depth are significantly higher than others, because of the similarity of \CO~and \COO~peak
temperatures.

For clumps at the boundary of the observed zone a flag ``X'' is
added to indicate that the quoted clump mass value actually is a
lower limit.

Comparing these results with the list of 29 dust cores presented in \citetalias{mas06b} and obtained applying
the 2D version of CF to 1.2~mm continuum observations of VMR-D at the SEST, we searched for possible
associations between \COO~and dust condensations, adopting as a criterium the correspondence between the dust
core centroid and one of the map points assigned by CF to a gas clump. We found 16 cases that are reported in
Table~\ref{clumptable}. Note that in this way the information on the velocity field, allowing to separate
possible multiple components along the line of sight, can induce the association between two gas clumps and a
group of dust cores (see, for example, the clumps VMRD36 and VMRD37). On the other hand, by comparing the two
lists in the opposite way, we can note that only the dust cores MMS10 and MMS11 are not associated to
\COO~clumps. This implies that, in general, the continuum emission is able to track even denser zones than
\COO~line, and in fact, considering the gas clumps associated to a group of \citetalias{mas06b} dust cores, we
find that the total mass is larger or quite similar to the mass traced by dust. The average value of their ratio
is in fact $M_{gas}$(line)$/M_{gas}$(continuum)$=1.44 \pm 0.11$, while the median of the distribution is 0.96.
Furthermore, the clumps that remain unassociated have intermediate or small masses ($M \leq 20$~M$_{\sun}$).
Finally, in Table~\ref{clumptable} the associations with the \textit{IRAS} PSC sources of Table~\ref{iraspsc}
are also indicated: 22 out of 25 objects fall in the area observed in \COO, and the locations of 15 sources
\citep[among which are the five IRS objects of][]{lis92} can be associated with the recognized clumps.

\begin{figure}
\epsscale{.7} \plotone{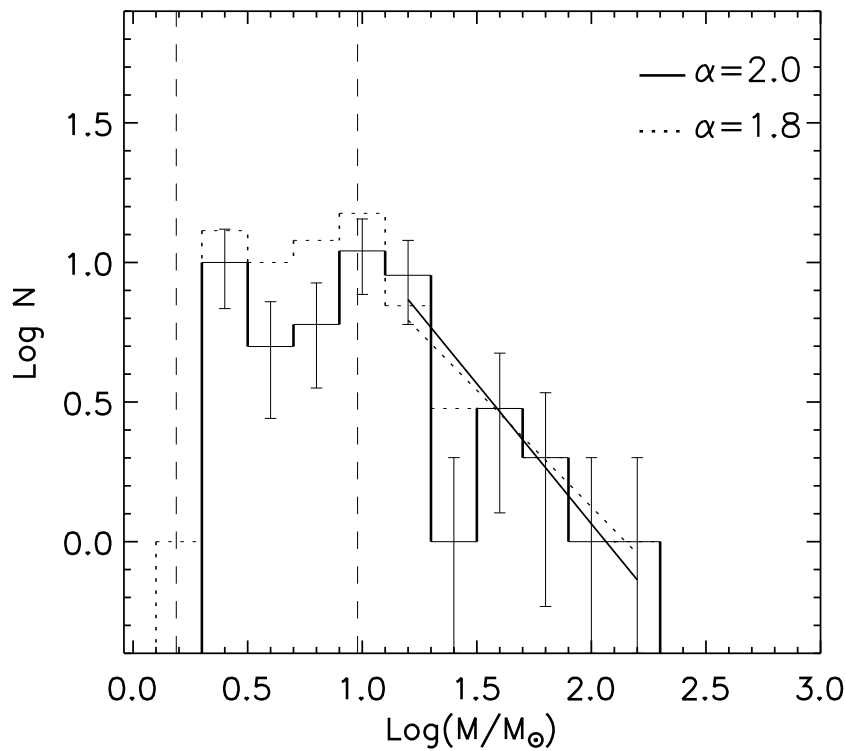} \caption{Clump mass spectrum of
VMR-D derived from \COO~observations using CF. The dashed vertical
lines indicate the minimum mass (leftmost) and the completeness
(rightmost) limits, respectively. The dotted line shows the CF mass
spectrum while the solid line is obtained after the merging
procedure (see text). The linear best-fit is also shown in both
cases, whose slope corresponds to $\alpha-1$, where $\alpha$ is the
exponent of the power law $\mathrm{d}N/\mathrm{d}M\propto
M^{-\alpha}$. \label{masscl13}}
\end{figure}

The derived mass spectrum is shown in Figure~\ref{masscl13}, with Poisson error bars, along with the spectrum
resulting from the original CF output. For nine clumps we can only give a lower limit to the mass because they
are bounded by our map limits. This effect is particularly important in the case of the large clump hosting
IRS~16 (clump VMRD1 in Table~\ref{clumptable}). The minimum mass detectable and the completeness limits,
calculated as in \citet{sim01} by setting a 10$\sigma$ confidence level \cite[see also][]{bai06}, and estimated
in 0.19~M$_{\sun}$ and 0.98~M$_{\sun}$, respectively, are also shown.

A linear least-squares fit of the distribution of the masses above
the completeness limit provides an estimate of the
$\mathrm{d}N/\mathrm{d}M\propto M^{-\alpha}$ law exponent, which has
been found to be $\alpha=2.0\pm 0.3$ for the ``corrected'' sample
and $\alpha=1.8\pm 0.3$ for the original CF output (see
Figure~\ref{masscl13}). In general, dealing with a relatively small
sample of clumps, the spectral index can be also sensitive to the
particular choice of the bin size. It was set here to be 0.2 (in
logarithmic units), after verifying the stability of the resulting
slope with respect to other possible choices. By varying, as in
\citet{bai06}, the bin size in the range of $\pm 10\%$ we obtain
slopes that always fall within the associated error.

Concluding this discussion about the \COO~clump mass distribution,
we remark that the observed values are significantly below those
calculated with the virial theorem: these two values often differ by
one order of magnitude, even considering different internal density
distributions. Interestingly, the same result is found for the
millimeter dust core distribution by \citetalias{mas06b}, who
indicate two possible confining mechanisms related either to
external pressure or to toroidal magnetic fields, excluding the
first one. The hypothesis that these clumps, or part of them, are
gravitationally bound cannot be excluded a priori if they are
collapsing, as could be suggested by the presence of associated IRAS
sources (see Table~\ref{clumptable}). However a particular trend in
the mass ratio cannot be recognized in these cases, so that it
remains difficult a definite conclusion about the collapsing or
dispersing status of these clumps.

In applying the CF algorithm to the \CO~line data cube, the large
number of blended components, as well as the presence of a diffuse
component in a large fraction of the map, make the decomposition
much more difficult and the definition of ``clump'' more uncertain.
Other difficulties are related to the optical depth of this line
that can be affected by self-absorption. However, the
\CO~observations are characterized by a better sampling with respect
to \COO, approximately 1 beam and 2 beams, respectively. In this
situation it is actually difficult to discern which line produces
the most reliable slope for the mass spectrum.

Even for the \CO~line the previous merging procedure was applied,
obtaining 168 objects from the original 275 ones. The cataloged
emission in clumps accounts for the 84\% of the total emission in
the \CO~map.

It is noteworthy that the positions of the \COO~clumps and the
\citetalias{mas06b} dust cores coincide with those of \CO~clumps,
suggesting that this line, although optically thick, can be anyway
used to reliably track the gas clumps.

\begin{figure}
\epsscale{0.7} \plotone{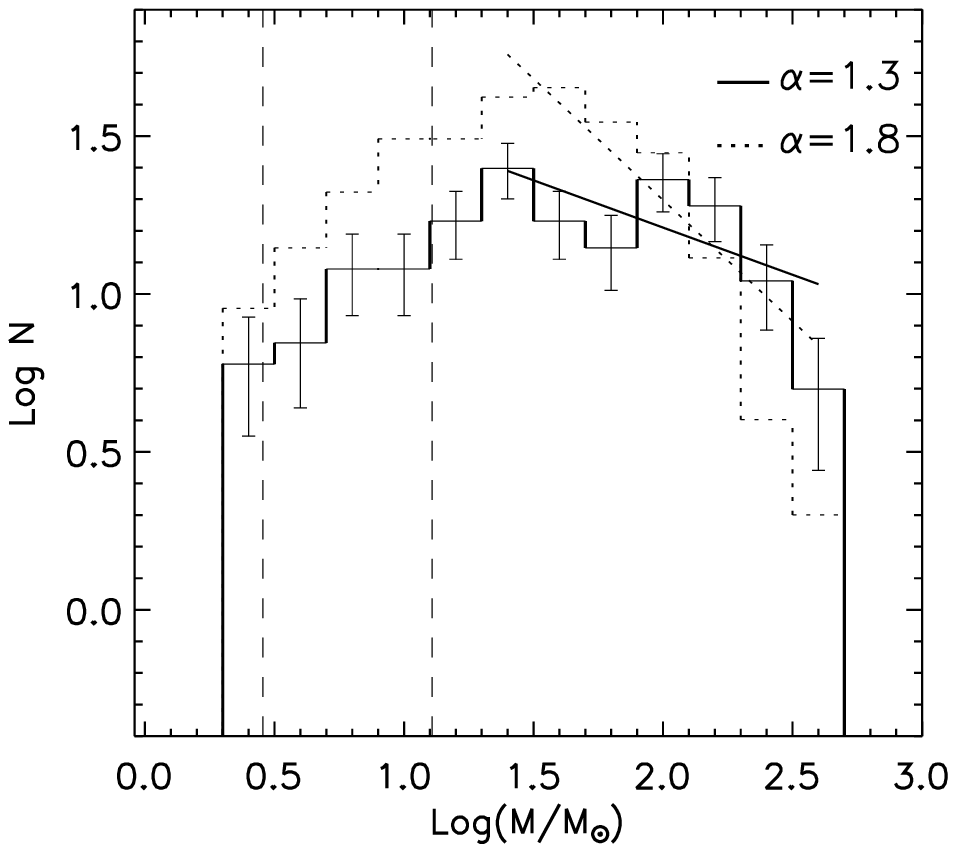} \caption{As in Figure
\ref{masscl13}, but for \CO. \label{masscl12}}
\end{figure}

The method used to determine the \CO~clump masses is based on
Equation~\ref{mass12}: even if this empirical relationship can be
questioned in the case of small-scale structures, it is the only way
to exploit the \CO~data for mass determinations and is often used to
derive the clump mass spectra down to the lower masses \citep[see,
e.g.,][]{hei98}. The resulting mass spectrum presented in
Figure~\ref{masscl12} shows, in the case of the merged sample, a
``flat'' central part and an approximately linear decrease in the
large mass limit of the distribution. The minimum mass and the
completeness limits are, respectively, 2.9~M$_{\sun}$ and
12.6~M$_{\sun}$. The spectral indexes obtained with and without
merging are significantly different, $\alpha=1.3\pm 0.1$ and
$\alpha=1.8\pm 0.1$, respectively. While the second value is
consistent with the corresponding index derived for \COO, those
obtained after merging are clearly different. It is plausible that
the high optical depth of \CO makes much more difficult to define
the actual borders of clumps.

Note that the spectral indexes found in VMR-D for the corrected
clump lists agree with those reported in the literature for
molecular line observations \citep[$1.3\leq\alpha\leq1.9$, see,
e.g.,][]{mac04}, and are smaller than the typical slopes of the
stellar IMF \citep[e.g., $\alpha \sim 2.35$ in][]{sal55}.

The slopes obtained here seem also comparable with the power-law
index $\alpha_{dust}=1.45\pm0.2$ reported in \citetalias{mas06b} if
we consider the quoted errors and the fact that the dust continuum
observations completely include those high-mass clumps which are
only partially contained in our map (in particular the massive clump
hosting IRS~16).

The difference found between the \COO~and the dust indexes could be
due to opacity effects that in the case of the more massive clumps
can be considerable. This corresponds to an underestimate of the
larger masses and then to a steeper slope in the mass spectrum. For
another similar case in which $\alpha_{gas}>\alpha_{dust}$, see
\citet{bai06}.

\begin{figure}
\epsscale{1.} \plotone{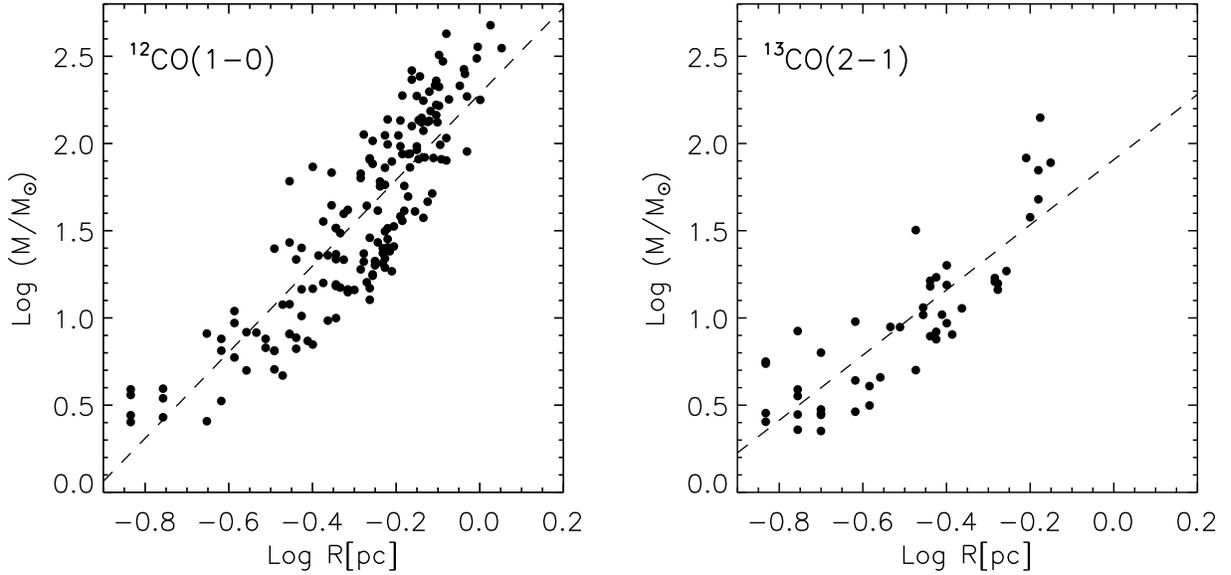} \caption{Bi-logarithmic plot of the
mass vs radius relation for the clumps resulting from the supervised
clump decomposition of VMR-D, for the \CO~and the \COO~transitions,
respectively. In both panels, the dashed line indicates the linear
trend of the data. The slopes are $x=2.5\pm 0.3$ for \CO, and
$x=1.9\pm 0.5$ for \COO. \label{massize}}
\end{figure}

Considering now the mass-radius relation shown in
Figure~\ref{massize} we can further extend the discussion in
\citetalias{mas06b} about the slope of this relation. In that work
the exponent of the $M\propto R^x$ relation has been evaluated
$x=1.7$, with this value decreasing when the masses are determined
by adopting two different temperature-opacity values for cores with
and without associated IRAS sources. Here on the other hand, with
reference to a weak linear behavior in the bi-logarithmic plot, we
find significantly different slopes in the case of \CO~and
\COO~clumps: $x=2.5\pm 0.3$ and $x=1.9\pm 0.5$, respectively. This
discrepancy can result from all the uncertainties involved in the
two derived mass distributions, in particular from the different
optical depths. The \COO~value appears more consistent with that of
\citetalias{mas06b}, both suggesting the presence of Bonnor-Ebert
clumps. Conversely, the slope derived from \CO~is more consistent
with a turbulent fragmentation scenario \citep[see, e.g.,][]{elm96}.

\begin{figure}
\epsscale{1.} \plotone{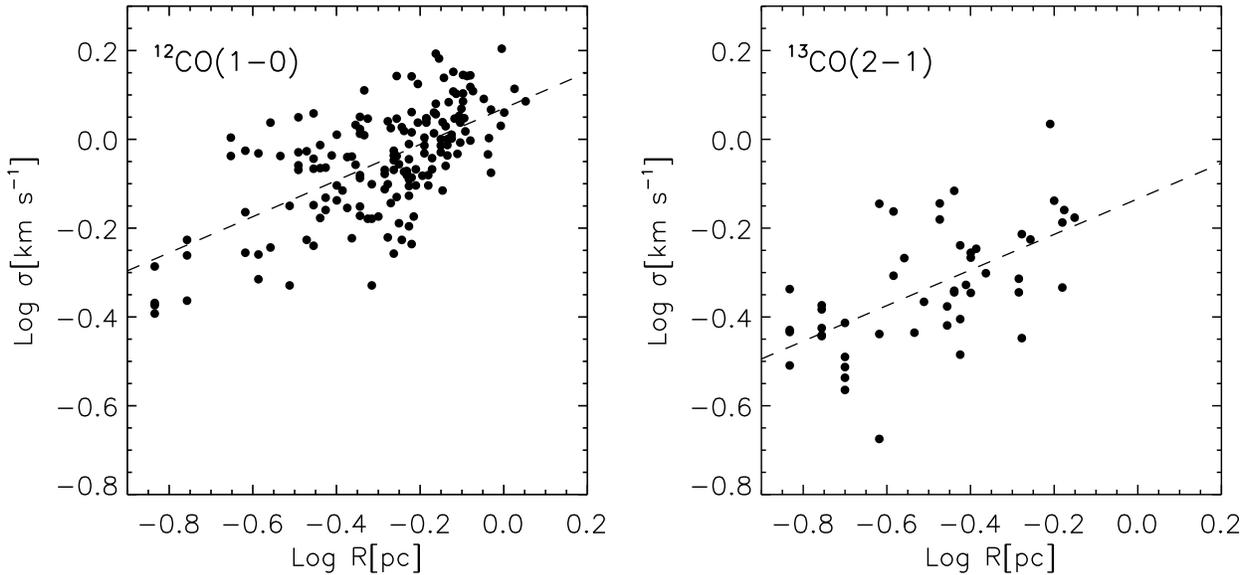} \caption{Bi-logarithmic plot of the
internal velocity dispersion vs radius for the VMR-D clumps. The
data are arranged as in Figure~\ref{massize}. The slopes are
$\beta=0.4\pm 0.1$ for \CO, and $\beta=0.4 \pm 0.2$ for
\COO.\label{larson}}
\end{figure}

Another relevant information resulting from the cloud decomposition
procedure is the internal velocity dispersion $\sigma$ of each clump
that can be used to examine its scaling relation with the radius $R$.
This is generally reported as $\sigma \propto R^{\beta}$
\citep[empirically derived by][]{lar81}, with $\beta$ varying
between 0.2 and 0.5 \citep[see, e.g.][]{sch04}, the most quoted
value being $\beta=0.4\pm0.1$ \citep[and references therein]{mac04}.
In Figure~\ref{larson} we present the scatter plot of the velocity
dispersion vs radius for the clumps observed in the two lines. If we
fit the data with a linear regression, the resulting slope is
$\beta=0.4\pm 0.1$ for \CO, and $\beta=0.4 \pm 0.2$ for \COO. These
values are not far from $\beta=0.5$, that is the value corresponding
to the virial equilibrium, a condition implying equipartition
between self-gravity and turbulent energies. The same equilibrium is
consistent with the slope $x=1.9$ we find for the mass vs radius
relation for the \COO.

On the other hand, other authors \citep[see, e.g.,][and references
therein]{mac04} consider a non-static scenario and explain this
value with the presence of supersonic turbulent cascades that appear
to be more consistent with the results we obtained above on the
gravitational stability of the \COO~clumps. Finally, it is
noteworthy that $\beta=0.5$ is also the value determined in two
surveys of the galactic plane carried out by \citet{dam86} and
\citet{sol87}.

\section{Summary} \label{summary}
In this work we examined the CO distribution in one of the most
interesting parts of the Vela Molecular Ridge (i.e., a $\sim 1\degr
\times 1\degr$ area of Cloud D) with unprecedented spatial
resolution ($\sim 50\arcsec$). The main reasons of interest for this
cloud are its star formation activity and its location in the
galactic plane. VMR-D is in fact the nearest massive star forming
region with this galactic position.

The millimeter line observations of the CO isotopes in VMR-D,
carried out at ESO-SEST, revealed a complex spatial and spectral
behavior down to the smallest spatial scales investigated ($\sim
1$~pc). The two rotational transitions analyzed, \CO~and \COO, due
to their different optical depth, allowed us to probe both low and
high density regions, confirming a tight link between the gas
distribution and the location of the brightest point-like far-IR
sources known in the field. The main filamentary structures,
appearing in the integrated intensity maps, show an arc-like shape
although it is not clear if this is produced by expanding shells or
stellar winds, both driven by nearby young (massive) stars. If an
expanding \textsc{Hii} region is considered, its dynamical age is
compatible with that estimated for the star formation activity.
Contributions from nearby known supernova remnants are excluded.
Another possibility is that the observed arcs are the product of the
internal cloud turbulent motions.

The analysis of the velocity field has shown several well-separated
zones, emitting at different velocities and suggesting a complex
structure and dynamics. In two regions, the presence of both two
velocity components (separated by $\sim 3.5$ and $\sim
4.5$~km~s$^{-1}$, respectively) and arc-shaped morphology in the
vel-pos diagrams can be interpreted as a signature of expanding
shells. These, compressing the gas, can be thought as the cause of
the observed cluster formation. The evolutionary times involved in
such a scenario are consistent with the star formation age of this
cloud. Further arc-like structures are also present in the vel-pos
diagrams which can be associated with the molecular arcs in the
integrated intensity maps.

The line profiles near the bright far-IR sources often appear
clearly broadened, suggesting the presence of outflows. A search for
such outflows in \CO~line near the \textit{IRAS} red sources shows
that, within the limits of resolution and noise of our observations,
there are 13 candidate objects worth of further and more detailed
observations for mapping their structure and accurately determining
their physical parameters.

One of the immediate developments of this work will be a
multi-wavelength analysis of this cloud with the aim of correlating
the clustered star formation \citep{lor93,mas00,mas06a} with the gas
and dust physical conditions.

Cloud decomposition has been carried out by means of the CLUMPFIND
code and subsequently corrected for spurious effects due to the
misinterpretation of the random noise fluctuations in the spectra.
We find that the clump mass spectra of \CO~and \COO~observations are
different. Both values are typical for interstellar molecular clouds
and, significantly, are in agreement with that obtained by
\citetalias{mas06b} for dust cores in VMR-D.

In all the investigated \COO~clumps the masses have been found to be
lower than the corresponding virial masses, revealing that these
clumps are either gravitationally unbound or collapsing.

A power-law fit of the clump mass vs radius shows that the exponent
are $x=2.5$ for \CO~and $x=1.9$ for \COO. Finally, a similar fit of
the internal velocity dispersion vs clump radius gives a slope
$\beta\sim 0.4$ in both transitions, a value suggesting either
virial equilibrium or internal turbulence. The first hypothesis is
however excluded when we take into account the inconsistency between
observed and virial clump masses.

\acknowledgments This work is partially supported by the Italian
Ministry for University and Research through the PRIN grants. We are
also grateful to the SEST staff for useful training, technical
assistance, and warm welcome. Finally we thank Indra Bains, Eric
Rosolowsky and Jonathan Williams who kindly gave us helpful
information about the cloud decomposition techniques.

{\it Facilities:} \facility{SEST ()}

\clearpage

\begin{deluxetable}{lcc}
\tabletypesize{\scriptsize} \tablecaption{Velocity ranges of the
main emitting regions in VMR-D\label{tabchann}}
\tablehead{
\colhead{Region} & \colhead{\CO} & \colhead{\COO}\\
                 & (km~s$^{-1}$)                   & (km~s$^{-1}$) \\
}
 \startdata
North-East & $-2\div12$   & $0\div4$, $6\div12$\\
North-West & $0\div6$, $10\div18$  & $4\div10$\\
Center     & $4\div16$ & $4\div12$ \\
West       & $0\div10$ & $0\div6$ \\
South (diffuse component) & $0\div6$ & \nodata \\
South-East & $4\div16$ & $6\div12$ \\
\enddata
\end{deluxetable}

\begin{deluxetable}{rlcccccrllrr}

\tabletypesize{\scriptsize}
\rotate
\tablecaption{\textit{IRAS} PSC sources
in the observed field, with fluxes increasing from 12~$\mu$m to
60~$\mu$m, and associated outflows.\label{iraspsc}}

\tablehead{\colhead{} & \colhead{\textit{IRAS}} &\colhead{$F_{12}$}&\colhead{$F_{25}$}&\colhead{$F_{60}$}&
\colhead{$F_{100}$}&\colhead{$L_{FIR}$}&\colhead{$V_{lsr}$}&\colhead{Outflow}& \colhead{Wing}&
\colhead{$V_\mathrm{range}$}&\colhead{Offset}\\
\colhead{} & \colhead{name}&\colhead{(Jy)}&\colhead{(Jy)}&\colhead{(Jy)}&\colhead{(Jy)}&\colhead{($L_\sun$)}&\colhead{(km
s$^{-1}$)} &\colhead{(Y/N/\ldots)} & &\colhead{(km s$^{-1}$)}
&\colhead{(\arcsec)}\\} \startdata
  1 & 08438-4340 (IRS 16)  &    13.4  &    56    &   638    &  1580    & 566 &   7.68 & Y    & blue\tablenotemark{\dagger} & 6.48 & $(0,-50)$       \\
    &                      &          &          &          &          &     &        &      & red              & 3.84 & $(+50,+50)$     \\
  2 & 08447-4309           & $<$ 0.25 & $<$ 0.4  & $<$ 4.4  &    36    &   6 &  11.28 & \ldots                                           \\
  3 & 08448-4343 (IRS 17)  &     8.7  &    88    &   327    &   1010   & 415 &   3.96 & Y    & blue             & 8.64 & $(0,0)$         \\
    &                      &          &          &          &          &     &        &      & red\tablenotemark{\dagger}  & 8.88 & $(-50,0)$       \\
  4 & 08448-4341           &     1.3  &     6.7  & $<$ 327  & $<$ 1010 & 258 &   4.20 & N                                                \\
  5 & 08453-4335           & $<$ 0.34 &     0.59 & $<$ 6.4  & $<$ 49   &   9 &   6.12 & N                                                \\
  6 & 08454-4307           & $<$ 0.25 &     0.36 & $<$ 3.3  & $<$ 39   &   6 &   2.52 & Y\tablenotemark{*}& blue             & 1.92  & (+100,+50)     \\
    &                      &          &          &          &          &     &        &      & red              & 6.72  & (+50,0)        \\
  7 & 08458-4332           &     1.1  &     2.7  &    18    &     53   &  20 &  12.00 & \ldots                                           \\
  8 & 08459-4338           & $<$ 0.29 &     0.57 &     6.5  &     38   &   8 &   6.00 & Y\tablenotemark{*}& blue\tablenotemark{\dagger} & 4.32  & (-100,0)       \\
  9 & 08461-4314           &     0.69 &     0.79 &     5.9  & $<$ 49   &   9 &   2.52 & \ldots                                           \\
 10 & 08463-4343           & $<$ 0.29 &     0.49 &     7.4  & $<$ 44   &   8 &  10.44 & Y\tablenotemark{*}& blue             & 3.48 & (+50,0)         \\
    &                      &          &          &          &          &     &        &      & red              & 6.24 & (+50,+100)      \\
 11 & 08465-4320           & $<$ 0.25 & $<$ 1.1  & $<$ 3.5  &     30   &   6 &  12.00 & \ldots                                           \\
 12 & 08468-4345           &     0.39 & $<$ 0.87 & $<$ 4.8  & $<$ 36   &   7 &   9.60 & Y\tablenotemark{*}& red\tablenotemark{\dagger}  & 6.12 & (+100,0)        \\
 13 & 08468-4330           & $<$ 0.3  &     0.43 &     3.4  & $<$ 32   &   5 &   6.72 & N                                                \\
 14 & 08470-4321 (IRS 19)  &    45    &   130    &   343    &    407   & 514 &  12.96 & Y    & blue\tablenotemark{\dagger} & 4.56 & (0,-50)         \\
    &                      &          &          &          &          &     &        &      & red              & 6.00 & (0,-50)         \\
 15 & 08471-4346           & $<$ 0.28 & $<$ 0.5  &     5.4  &     36   &   7 &   8.16 & \ldots                                           \\
 16 & 08472-4326A          &     0.85 &     0.94 &    11    & $<$ 407  &  45 &  12.48 & N                                                \\
 17 & 08474-4325           & $<$ 0.3  &     0.96 & $<$ 16   &     58   &  15 &  13.20 & Y\tablenotemark{*}& blue\tablenotemark{\dagger} & 8.16 & (0,0)           \\
    &                      &          &          &          &          &     &        &      & red              & 4.68 & (0,+50)         \\
 18 & 08475-4352           & $<$ 0.44 & $<$ 0.45 & $<$ 2.9  &     31   &   5 &   8.52 & \ldots                                           \\
 19 & 08476-4306 (IRS 20)  &     5.7  &     44   &   216    &    504   & 234 &   2.28 & Y\tablenotemark{*}& blue             & 2.64 & (0,0)           \\
    &                      &          &          &          &          &     &        &      & red              & 4.68 & (0,-50)         \\
 20 & 08477-4359 (IRS 21)  &     9    &     26   &   317    &    581   & 265 &   7.80 & Y\tablenotemark{*}& blue\tablenotemark{\dagger} & 5.88 & (+50,-50)       \\
    &                      &          &          &          &          &     &        &      & red\tablenotemark{\dagger}  & 4.44 & (0,-50)         \\
 21 & 08478-4403           &     0.46 &     0.51 &     5.4  & $<$ 581  &  56 &  10.20 & \ldots                                           \\
 22 & 08478-4303           & $<$ 0.39 &     1.6  & $<$ 216  & $<$ 35   & 108 &   4.92 & N                                                \\
 23 & 08479-4311           & $<$ 0.32 &     0.4  & $<$ 8.7  & $<$ 38   &   8 &   2.52 & Y\tablenotemark{*}& blue             & 3.60 & (0,+50)         \\
    &                      &          &          &          &          &     &        &      & red              & 2.52 & (+50,+50)       \\
 24 & 08483-4305           &     1.5  &     2.5  &  $<$ 39  &   189    &  42 &   3.36 & Y\tablenotemark{*}& blue             & 7.80 & (+100,0)        \\
    &                      &          &          &          &          &     &        &      & red              & 2.04 & (+50,+50)       \\
 25 & 08484-4350           &     0.33 & $<$ 0.56 & $<$ 3.7  &    23    &   5 &   6.24 & Y\tablenotemark{*}& blue             & 1.80 & (-50,0)         \\
    &                      &          &          &          &          &     &        &      & red              & 3.24 & (-100,0)        \\
\enddata
\tablenotetext{*}{A second component present in the spectrum has
been preliminarily fitted and subtracted.}

\tablenotetext{\dagger}{The wing profile is probably contaminated by
the contribution of very close velocity components.}
\end{deluxetable}

\begin{deluxetable}{lrrrrrrrrrcll}
\tabletypesize{\scriptsize}
\rotate
\tablecaption{List of the clumps
detected in the \COO~line.\label{clumptable}} \tablewidth{500pt}
\tablehead{\colhead{} & \colhead{$\alpha$ (2000)} &
\colhead{$\delta$ (2000)} & \colhead{$V_{lsr}$} & \colhead{$\Delta V$} & \colhead{Radius}&
\colhead{Mass}& \colhead{$M_{vir}$}&  \colhead{$\tau_{V,MAX}$}&
\colhead{$\int\tau_V\,dV$} & \colhead{Map limit}&
\colhead{Dust}&\colhead{IRAS}\\
\colhead{} & \colhead{h~~m~~s}&\colhead{\degr~~\arcmin~~\arcsec}&
\colhead{(km s$^{-1}$)} &\colhead{(km s$^{-1}$)}&\colhead{(pc)}& \colhead{(M$_{\sun}$)}&
\colhead{(M$_{\sun}$)}& \colhead{(km s$^{-1}$)$^{-1}$}&\colhead{}&\colhead{}&
\colhead{clump(s)}& \colhead{source(s)\tablenotemark{a}}\\}
\startdata
 VMRD1 & 8  45  35 &  -43  52  02 &  5.9 & 2.55 & 0.6 &  83 &  842 & 0.90 & 2.58 & X &                  MMS1,2,3 &         1 \\
 VMRD2 & 8  45  36 &  -43  48  42 &  5.3 & 1.80 & 0.4 &  16 &  248 & 0.62 & 0.66 & X &                  MMS1,2,3 &         1 \\
 VMRD3 & 8  46  17 &  -43  56  12 &  4.1 & 1.14 & 0.5 &  17 &  142 & 0.26 & 0.34 &   &                   \nodata &   \nodata \\
 VMRD4 & 8  46  31 &  -43  54  32 &  2.9 & 1.09 & 0.7 &  48 &  165 & 0.69 & 0.75 &   &                      MMS4 &      3,4 \\
 VMRD5 & 8  46  35 &  -43  54  32 &  5.0 & 1.53 & 0.7 &  70 &  324 & 0.97 & 2.39 &   &                  MMS4,5,6 &      3,4 \\
 VMRD6 & 8  46  49 &  -43  50  22 &  4.7 & 1.06 & 0.4 &   9 &   94 & 0.26 & 0.06 &   &                   \nodata &   \nodata\\
 VMRD7 & 8  46  54 &  -43  54  32 &  4.1 & 1.71 & 0.6 &  38 &  388 & 0.33 & 0.17 &   &                    MMS5,6 &   \nodata \\
 VMRD8 & 8  47  26 &  -43  48  42 &  5.6 & 1.11 & 0.4 &  10 &   99 & 0.32 & 0.46 &   &                   \nodata &   \nodata \\
 VMRD9 & 8  47  26 &  -43  52  52 &  2.9 & 1.36 & 0.4 &  17 &  145 & 0.45 & 0.44 &   &                   \nodata &   \nodata \\
VMRD10 & 8  47  31 &  -43  52  52 &  4.1 & 0.93 & 0.4 &   8 &   67 & 0.41 & 0.39 &   &                   \nodata &   \nodata \\
VMRD11 & 8  47  40 &  -43  43  42 & 11.3 & 0.99 & 0.3 &  11 &   72 & 0.59 & 0.50 &   &                   \nodata &         7 \\
VMRD12 & 8  47  45 &  -43  47  02 &  6.8 & 0.84 & 0.5 &  15 &   78 & 0.28 & 0.22 &   &                   \nodata &   \nodata \\
VMRD13 & 8  47  45 &  -43  50  22 &  5.0 & 1.31 & 0.4 &  15 &  143 & 0.42 & 0.71 &   &                   \nodata &         8 \\
VMRD14 & 8  47  50 &  -43  47  52 &  5.9 & 0.77 & 0.4 &   8 &   46 & 0.22 & 0.21 &   &                   \nodata &         8 \\
VMRD15 & 8  47  54 &  -43  27  52 &  2.0 & 0.73 & 0.2 &   3 &   16 & 0.34 & 0.42 & X &                   \nodata &   \nodata \\
VMRD16 & 8  47  59 &  -43  39  32 & 10.4 & 0.64 & 0.2 &   2 &   17 & 0.22 & 0.12 &   &                      MMS7 &   \nodata \\
VMRD17 & 8  47  59 &  -43  25  22 &  2.3 & 0.68 & 0.2 &   3 &   19 & 0.62 & 0.20 & X &                   \nodata &         9 \\
VMRD18 & 8  47  59 &  -43  20  22 &  1.1 & 0.50 & 0.2 &   3 &   12 & 0.78 & 0.60 & X &                   \nodata &   \nodata \\
VMRD19 & 8  48  03 &  -43  51  12 &  4.7 & 0.76 & 0.2 &   3 &   24 & 2.58 & 1.44 &   &                   \nodata &   \nodata \\
VMRD20 & 8  48  13 &  -43  47  02 &  7.7 & 1.62 & 0.3 &   3 &  143 & 0.28 & 0.16 &   &                   \nodata &   \nodata \\
VMRD21 & 8  48  13 &  -43  22  52 &  2.3 & 1.07 & 0.4 &  15 &   86 & 1.03 & 0.73 &   &                   \nodata &   \nodata \\
VMRD22 & 8  48  17 &  -43  17  52 &  1.4 & 1.08 & 0.2 &   5 &   36 & 0.62 & 0.62 &   &                   \nodata &   \nodata \\
VMRD23 & 8  48  17 &  -43  26  12 &  2.3 & 0.88 & 0.2 &   3 &   23 & 0.57 & 0.65 &   &                   \nodata &   \nodata \\
VMRD24 & 8  48  27 &  -43  41  12 &  6.5 & 1.69 & 0.3 &   5 &  201 & 0.48 & 0.11 &   &                   \nodata &   \nodata \\
VMRD25 & 8  48  31 &  -43  22  02 &  2.0 & 0.85 & 0.2 &   4 &   26 & 0.54 & 0.44 &   &                   \nodata &   \nodata \\
VMRD26 & 8  48  31 &  -43  25  22 &  2.3 & 0.91 & 0.2 &   6 &   34 & 0.42 & 0.53 &   &                   \nodata &   \nodata \\
VMRD27 & 8  48  31 &  -43  34  32 & 12.5 & 1.18 & 0.4 &  11 &  125 & 0.47 & 0.24 &   &                   \nodata &        11 \\
VMRD28 & 8  48  40 &  -43  29  32 & 11.3 & 1.33 & 0.4 &   8 &  153 & 0.50 & 0.45 &   &        MMS8,9,12,14,15,16 &        14 \\
VMRD29 & 8  48  40 &  -43  17  52 &  2.3 & 0.87 & 0.2 &   6 &   23 & 0.88 & 0.53 &   &                   \nodata &   \nodata \\
VMRD30 & 8  48  50 &  -43  29  32 & 13.1 & 1.07 & 0.5 &  16 &  123 & 0.39 & 0.20 &   &     MMS8,9,12,13,14,15,16 &        14 \\
VMRD31 & 8  48  54 &  -43  15  22 &  2.0 & 0.97 & 0.2 &   8 &   35 & 0.79 & 0.78 &   &                   \nodata &   \nodata \\
VMRD32 & 8  48  59 &  -43  59  32 &  8.3 & 1.44 & 0.5 &  16 &  229 & 0.81 & 0.53 &   &                   \nodata &        15 \\
VMRD33 & 8  49  03 &  -43  32  52 & 11.9 & 0.86 & 0.2 &   4 &   37 & 0.58 & 0.54 &   &                   \nodata &   \nodata \\
VMRD34 & 8  49  03 &  -43  37  52 & 11.9 & 1.69 & 0.2 &  10 &  143 & 0.54 & 0.71 &   &                  MMS17,18 &        16 \\
VMRD35 & 8  49  12 &  -43  22  02 &  1.4 & 1.27 & 0.3 &   5 &   93 & 0.37 & 0.52 &   &                   \nodata &   \nodata \\
VMRD36 & 8  49  13 &  -43  36  12 & 11.3 & 1.55 & 0.3 &  32 &  170 & 0.94 & 1.81 &   &               MMS19,20,21 &        17 \\
VMRD37 & 8  49  13 &  -43  36  12 & 12.8 & 0.86 & 0.3 &   9 &   45 & 0.64 & 0.46 &   &               MMS19,20,21 &        17 \\
VMRD38 & 8  49  22 &  -44  00  22 &  8.3 & 1.40 & 0.6 &  19 &  228 & 0.39 & 0.24 &   &                   \nodata &   \nodata \\
VMRD39 & 8  49  22 &  -43  58  42 &  7.4 & 1.07 & 0.4 &   8 &   88 & 1.01 & 0.81 &   &                   \nodata &   \nodata \\
VMRD40 & 8  49  26 &  -43  22  02 &  3.5 & 0.72 & 0.2 &   3 &   21 & 0.32 & 0.30 &   &                   \nodata &   \nodata \\
VMRD41 & 8  49  26 &  -43  17  02 &  2.6 & 0.90 & 0.3 &  10 &   59 & 0.28 & 0.37 &   &                  MMS22,24 &        19 \\
VMRD42 & 8  49  26 &  -43  17  02 &  1.7 & 0.85 & 0.2 &   3 &   26 & 0.26 & 0.23 &   &                  MMS22,24 &        19 \\
VMRD43 & 8  49  27 &  -44  10  22 &  8.9 & 1.57 & 0.7 &  78 &  365 & 3.55 & 5.50 &   &               MMS25,26,27 &        20 \\
VMRD44 & 8  49  31 &  -43  20  22 &  7.7 & 1.16 & 0.3 &   4 &   73 & 0.40 & 0.42 &   &                   \nodata &   \nodata \\
VMRD45 & 8  49  31 &  -44  07  52 &  9.8 & 1.63 & 0.7 & 141 &  373 & 1.10 & 0.59 &   &            MMS23,25,26,27 &        20 \\
VMRD46 & 8  49  49 &  -43  21  12 &  8.3 & 1.01 & 0.3 &   9 &   66 & 0.63 & 0.69 &   &                   \nodata &        23 \\
VMRD47 & 8  49  59 &  -44  00  22 &  6.8 & 1.00 & 0.2 &   2 &   36 & 0.32 & 0.30 &   &                   \nodata &   \nodata \\
VMRD48 & 8  50  03 &  -43  22  52 &  9.5 & 1.27 & 0.4 &  20 &  136 & 0.60 & 0.81 & X &                   \nodata &   \nodata \\
VMRD49 & 8  50  07 &  -43  16  12 &  4.1 & 0.88 & 0.2 &   4 &   28 & 0.35 & 0.30 & X &                     MMS28 &        24 \\

\enddata
\tablenotetext{a}{The identification numbers are the same as in
Table~\ref{iraspsc}.}
\end{deluxetable}

\clearpage

\end{document}